\documentclass[sigconf]{acmart}

\AtBeginDocument{%
  }

\setcopyright{none}
\copyrightyear{2026}
\acmYear{2026}
\acmDOI{XXXXXXX.XXXXXXX}
\acmConference[ '26]{XX ( )}{XX }{XX}
\acmISBN{978-1-4503-XX-X/26/XX}
\renewcommand\footnotetextcopyrightpermission[1]{}

\usepackage{booktabs}
\usepackage{multirow}

\usepackage{amsmath,amssymb}
\usepackage{enumitem}
\usepackage{xcolor}
\usepackage{url}
\usepackage{xspace}
\usepackage[switch]{lineno}
\usepackage{graphicx}
\usepackage{pifont}
\usepackage{listings}

\usepackage{tabularx}
\newcolumntype{Y}{>{\raggedright\arraybackslash}X}

\newcommand{\benchname}{\textsc{SysTradeBench}\xspace}
\newcommand{\benchabbr}{\textsc{SysTB}\xspace}

\newcommand{\cmark}{\textcolor{green!60!black}{\ding{51}}}
\newcommand{\xmark}{\textcolor{red!70!black}{\ding{55}}}
\newcommand{\pmark}{\textcolor{gray!70}{\(\circ\)}}

\newcommand{\PlaceFigure}[2]{%
  \IfFileExists{#1}{%
    \includegraphics[width=\linewidth]{#1}%
  }{%
    \fbox{%
      \parbox{0.97\linewidth}{%
        \centering\vspace{1.5em}
        \textbf{Placeholder for Figure: #2}\\
        \vspace{0.5em}
        \texttt{#1}\\
        \vspace{1.5em}
      }%
    }%
  }%
}

\begin{document}

\title{\benchname: An Iterative Build--Test--Patch Benchmark for Strategy-to-Code Trading Systems with Drift-Aware Diagnostics}

%% Authors
\author{Yuchen CAO}
\email{cynthcao2-c@my.cityu.edu.hk}
\affiliation{%
  \institution{City University of Hong Kong}
  \city{Hong Kong SAR}
  \country{China}
}

\author{Hanlin Zhang}
\email{hanlzhang8-c@my.cityu.edu.hk}
\affiliation{%
  \institution{City University of Hong Kong}
  \city{Hong Kong SAR}
  \country{China}
}

\author{Jacky Wai Keung}
\email{Jacky.Keung@cityu.edu.hk}
\affiliation{%
  \institution{City University of Hong Kong}
  \city{Hong Kong SAR}
  \country{China}
}

\author{Yang Chen}
\authornote{Corresponding author. \newline Code repository: \url{https://github.com/YgcCoder/SysTB}}
\email{ychen2014@zju.edu.cn}
\affiliation{%
  \institution{Zhejiang University}
  \city{Hangzhou}
  \country{China}
}

\author{Linqi SONG}
\email{linqi.song@cityu.edu.hk}
\affiliation{%
  \institution{City University of Hong Kong}
  \city{Hong Kong SAR}
  \country{China}
}

\renewcommand{\shortauthors}{CAO et al.}

%% Abstract
\begin{abstract}
Large Language Models (LLMs) are increasingly used as quantitative research copilots to translate natural-language strategy specifications into executable trading code.
Yet most existing evaluations either focus on static financial knowledge or summarize performance with a single profitability metric, leaving a gap for benchmarking \emph{strategy-to-code} trading systems as governed, auditable software.
We introduce \benchname (\benchabbr), an iterative \emph{build--test--patch} benchmark that evaluates LLM-generated trading systems under drift-aware diagnostics.
Given a standardized Base Strategy Doc and frozen semantics, each model must produce (i) a strategy card, (ii) executable code, and (iii) mandatory audit logs.
A sandboxed harness runs determinism and anti-leakage checks, detects rule drift across iterations, and returns evidence bundles to support constrained patches.
\benchname reports multi-dimensional scorecards (D1: Spec Fidelity, D2: Risk Discipline, D3: Reliability, D4: OOS Robustness Indicators) and cost--effectiveness signals.
We evaluate 17 models across 12 strategies: top models achieve $\geq$91.7\% validity with 7.29--7.85 scores, but evidence-driven iteration induces code convergence (95.4\% similarity by Iter2). These findings suggest that LLM iteration complements rather than replaces human Quantitative Researcher (QR) governance: LLMs excel at rapid prototyping and shallow bug fixes, while QR oversight remains essential for critical strategies requiring solution diversity and ensemble robustness.
\end{abstract}



\maketitle

%% ============================================================
\section{Introduction}
\label{sec:intro}

\noindent Large Language Models (LLMs) have evolved from passive text assistants into tool-using systems that can retrieve information, write executable code, and perform multi-step decision making~\cite{vaswani2017attention,brown2020language,ouyang2022instructgpt,achiam2023gpt4,yao2022react,schick2023toolformer}.
This shift is particularly visible in software tasks: modern LLMs can generate code, follow API contracts, and iteratively debug with test feedback~\cite{chen2021evalcode,austin2021mbpp,li2022alphacode,jimenez2023swebench}.
In practice, many deployments combine instruction following with retrieval and tool augmentation (e.g., retrieval-augmented generation) to ground decisions and reduce hallucinations~\cite{lewis2020rag,liang2022helm}.

\noindent In quantitative finance, these capabilities enable a new workflow: analysts describe a rule-based trading idea in natural language, and the model generates runnable strategy code that can be backtested and refined.
However, \emph{strategy-to-code} systems are not merely prediction models.
They are \emph{governed software} that must implement intended semantics (fidelity), satisfy operational constraints (risk and session rules), execute deterministically, avoid information leakage, and produce audit-grade traces that connect signals$\rightarrow$risk checks$\rightarrow$orders$\rightarrow$positions$\rightarrow$P\&L.
Such requirements mirror institutional practice in systematic trading where engineering controls, reproducibility, and auditability are as critical as raw returns~\cite{lopezdeprado2018afml,keim1997costs,almgren2001optimal,artzner1999coherent}.

\noindent A single headline backtest metric (e.g., Sharpe) can be misleading for at least three reasons.
First, net performance is often dominated by turnover and trading frictions rather than gross signal quality~\cite{barber2000trading,keim1997costs,frazzini2015tradingcosts,hasbrouck2007microstructure}.

Second, backtests are vulnerable to data-snooping and selection bias; ``best-in-sample'' strategies may fail out of sample without overfitting-aware validation~\cite{white2000reality,sullivan1999datasnooping,bailey2017pbo,harvey2016pvalues,lo2002sharpe}.
Third, subtle leakage and non-determinism can inflate results while being difficult to detect without system-level diagnostics and strict execution harnesses.

\begin{table}[!b]
\centering
\small
\setlength{\tabcolsep}{3.8pt}
\renewcommand{\arraystretch}{1.05}
\begin{tabular}{@{}lcccccc@{}}
\toprule
\textbf{Benchmark} & \textbf{S$\rightarrow$C} & \textbf{Exec} & \textbf{Iter} & \textbf{Drift} & \textbf{Det/Leak} & \textbf{Audit} \\
\midrule
Market-Bench          & \cmark & \cmark & \xmark & \xmark & \pmark & \xmark \\
QuantEval             & \pmark & \cmark & \xmark & \xmark & \pmark & \pmark \\
StockBench            & \xmark & \cmark & \xmark & \xmark & \pmark & \xmark \\
LiveTradeBench        & \xmark & \cmark & \xmark & \xmark & \cmark & \xmark \\
PredictionMarketBench & \xmark & \cmark & \pmark & \xmark & \cmark & \pmark \\
\midrule
\benchabbr\ (this work) & \cmark & \cmark & \cmark & \cmark & \cmark & \cmark \\
\bottomrule
\end{tabular}
\caption{Capability coverage across representative benchmarks.
\cmark\ denotes first-class support with an explicit task interface and reproducible evaluation;
\xmark\ denotes not supported;
\pmark\ denotes partial/optional/implicit coverage (mentioned or enabled in some settings, but not systematically required nor evaluated as a core criterion).
S$\rightarrow$C: specification-to-code contract; Exec: sandboxed execution; Iter: build--test--patch; Drift: drift-aware diagnostics; Det/Leak: determinism and anti-leakage checks; Audit: mandatory structured audit logs.}
\label{tab:gaps}
\end{table}

\noindent \textbf{Positioning w.r.t.\ prior work.}
Existing benchmarks contribute complementary pieces.
For example, Market-Bench emphasizes numerically verifiable backtester construction from strategy descriptions~\cite{srivastava2025marketbench}, and QuantEval evaluates strategy coding alongside quantitative reasoning and financial QA under reproducible configurations~\cite{kang2026quanteval}.
Agent benchmarks such as StockBench and LiveTradeBench study sequential decision-making under contamination-aware or live settings~\cite{chen2025stockbench,yu2025livetradebench}, and PredictionMarketBench explores replay-style deterministic evaluation for event-driven markets~\cite{arora2026predictionmarketbench}.
Our focus is different: we benchmark \emph{system-level correctness and governability} (fidelity, constraint discipline, determinism, leakage resistance, auditability) and explicitly measure \emph{iterative repair} under constrained patch budgets while penalizing semantic drift---a failure mode analogous to ``changing the task'' rather than fixing the implementation.

\noindent \textbf{Capability coverage across representative benchmarks.}
Table~\ref{tab:gaps} summarizes representative benchmarks and the capabilities they emphasize.
Rather than treating any single benchmark as ``complete'', \benchname targets evaluation signals most relevant to deploying LLM-generated strategy code as auditable systems: (i) strict spec-to-code contracts, (ii) iterative build--test--patch with evidence bundles, (iii) drift-aware diagnostics, and (iv) mandatory audit logs alongside deterministic, leakage-free execution.

\noindent \textbf{\benchname: iterative build--test--patch benchmarking with drift-aware diagnostics.}
We introduce \benchname (\benchabbr), a system-centric benchmark that evaluates \emph{strategy-to-code} trading systems as governed software artifacts rather than as one-shot profit generators.
Given a standardized Base Strategy Doc with frozen, machine-checkable semantics, a model must produce three auditable artifacts: (i) a structured strategy interpretation (a \emph{strategy card}), (ii) runnable modular code, and (iii) mandatory structured logs that trace \texttt{signal$\rightarrow$risk\_check$\rightarrow$order$\rightarrow$position$\rightarrow$P\&L}.
A sandboxed harness executes submissions under fixed seeds and frozen data splits, validates determinism and temporal integrity (anti-leakage), stress-tests constraint compliance, and returns an evidence bundle (metrics, violations, audit gaps, and test reports).
Crucially, \benchname supports \emph{controlled patching}: models iteratively submit constrained repairs based on evidence bundles, while drift-aware diagnostics detect and penalize silent semantic drift so that improvements reflect implementation quality rather than strategy changes.

\noindent \textbf{Contributions.}We make the following contributions:
\begin{itemize}[leftmargin=*]
  \item \textbf{Governed strategy-to-code benchmark interface:} Base Strategy Docs with frozen semantics and mandatory output contracts (strategy card, executable code, and audit logs) across 12 canonical strategies.
  \item \textbf{Evidence-driven build--test--patch protocol:} iterative repair under explicit patch budgets, enabling trajectory analysis beyond one-shot generation.
  \item \textbf{Drift-aware system diagnostics:} checksum- and trace-based checks that penalize unauthorized semantic drift across iterations.
  \item \textbf{Multi-dimensional scorecards with transaction-cost-aware framework:} system-level evaluation of fidelity, risk governance, reliability/auditability, and OOS robustness signals (preliminary evaluation on sampled data), together with token/cost accounting for practical model selection.
  \item \textbf{Empirical characterization of evidence-driven repair dynamics:} we demonstrate that LLM iteration induces code convergence (95.4\% similarity by Iter2, byte-identical by Iter3) while achieving substantial quality gains, revealing a fundamental tension between automated efficiency and solution diversity that positions LLM repair as complementary to---rather than replacement for---human Quantitative Researcher governance.
\end{itemize}

%% ============================================================
\section{Related Work}
\label{sec:related}

\subsection{Financial-domain LLMs and data-centric adaptation}
\label{sec:rw_finllm}
\noindent Finance-specific adaptation ranges from encoder-style models such as FinBERT~\cite{araci2019finbert} to large generative models trained on curated financial corpora (e.g., BloombergGPT, FinGPT)~\cite{wu2023bloomberggpt,yang2023fingpt}. While these models improve finance-language understanding, they do not directly evaluate whether an LLM can generate \emph{auditable, governed strategy-to-code trading software} that remains stable under iterative debugging and strict constraints.

\subsection{Static finance benchmarks for QA and reasoning}
\label{sec:rw_static}
\noindent Static evaluation suites such as FinanceBench/FinBen, PIXIU, FLUE/FLANG, and numerical-report reasoning benchmarks (FinQA, TAT-QA) cover broad financial QA, instruction following, and table-text reasoning~\cite{islam2023financebench,xie2024finben,xie2023pixiu,shah2022flue,chen2021finqa,zhu2021tatqa}. However, their task interfaces typically do not require executable trading systems with deterministic harness execution, constraint enforcement, and audit-grade traces.

\subsection{Code generation, program repair, and iterative evaluation}
\label{sec:rw_code}
\noindent \benchname relates to test-based code generation and iterative repair: HumanEval/MBPP and AlphaCode evaluate functional correctness of generated programs~\cite{chen2021evalcode,austin2021mbpp,li2022alphacode}, while SWE-bench and program-repair systems (e.g., GenProg, SapFix) emphasize fixing issues under feedback and regression constraints~\cite{jimenez2023swebench,legoues2012genprog,marginean2019sapfix}. We adopt this build--test--patch philosophy but specialize it to trading governance: patches must preserve frozen semantics (no drift), executions must be deterministic/leakage-free, and artifacts must emit audit logs for traceability.

\subsection{Executable trading benchmarks and sequential decision making}
\label{sec:rw_exec}
\noindent Executable finance benchmarks provide complementary components: Market-Bench focuses on constructing executable backtesters from strategy descriptions~\cite{srivastava2025marketbench}, and QuantEval combines financial QA/reasoning with strategy execution under controlled settings~\cite{kang2026quanteval}. Agent-style environments (StockBench, LiveTradeBench, PredictionMarketBench) study sequential decision making under contamination-aware/live/replay settings, alongside tool-augmented agent systems such as FinMem and FinAgent~\cite{chen2025stockbench,yu2025livetradebench,arora2026predictionmarketbench,yu2023finmem,zhang2024finagent}. In contrast, \benchname targets \emph{system-level governability} for strategy-to-code artifacts (fidelity, constraint discipline, determinism/leakage resistance, and auditability) and measures evidence-driven repair trajectories under drift-aware diagnostics.

\subsection{Backtest overfitting, frictions, and statistical validity}
\label{sec:rw_stats}
\noindent Robust evaluation must account for data-snooping, multiple testing, and frictions: classic procedures include White's Reality Check and SPA-style tests~\cite{white2000reality,sullivan1999datasnooping,hansen2005spa}; robustness metrics include the Deflated Sharpe Ratio and Probability of Backtest Overfitting~\cite{bailey2014deflated,bailey2017pbo}; and empirical evidence shows trading costs can materially erode returns~\cite{keim1997costs,frazzini2015tradingcosts}. \benchname incorporates these ideas in D4 as robustness/diagnostic signals rather than profit-only objectives, consistent with recent findings that LLM investing gains weaken under stricter controls~\cite{li2025finsaber}.

\section{Illustrative Example: When ``Understanding'' Diverges from Code Quality}
\label{sec:illustrative_example}

We illustrate why \benchname evaluates \emph{strategy-to-code} systems as governed software rather than relying on a single performance metric or explanation quality. Using the \emph{Double Moving Average Crossover} strategy (go long on a golden cross, go short on a death cross, reverse only on the opposite signal), the Base Strategy Doc freezes key governance constraints (no look-ahead, SMA-only indicators, $N_{\text{short}}<N_{\text{long}}$, bar-close execution) and mandates structured outputs (a strategy card plus \texttt{trade\_log} and \texttt{audit\_log}) so that decisions are traceable from signals to orders and P\&L.

Despite identical inputs, models translate the same specification into code with markedly different implementation quality (Figure~\ref{fig:first}). Rankings based on articulation or LLM-as-judge impressions can diverge from execution-based system evaluation: some seemingly correct explanations correspond to fragile, non-auditable, or leakage-prone code, while concise implementations may score lower under explanation-focused judging yet behave more reliably in sandboxed checks. Failures include silent semantic drift during bug fixes, incomplete audit logs despite passing backtests, and look-ahead via \texttt{df.shift(-1)}. This motivates \benchname's design: hard gates (executability, determinism, anti-leakage, audit completeness) plus multi-dimensional scorecards and evidence-driven build--test--patch iterations that measure whether models can repair failures without changing frozen strategy semantics.

%% ============================================================
\section{SysTradeBench: Benchmark Design}
\label{sec:design}

\noindent This section details \benchname's three-layer architecture with iterative execution control (Figure~\ref{fig:second}). The workflow comprises: (1) \textbf{Input Layer} (§\ref{sec:input_layer}) provides standardized specification contracts, frozen semantics, market data, and evidence bundles; (2) \textbf{Model Generation Layer} (§\ref{sec:model_layer}) handles LLM-based code synthesis producing strategy cards, executable code, and audit logs; (3) \textbf{Evaluation Layer} (§\ref{sec:eval_layer}) performs multi-dimensional scoring (D1--D4), LLM cross-evaluation, and optional human testing; and (4) \textbf{Executor} (§\ref{sec:executor}) orchestrates sandboxed execution, drift detection, and iterative refinement.

%% ============================================================
\subsection{System Architecture and Workflow}
\label{sec:arch}

\label{sec:arch}

\begin{figure*}[t]
\centering
\includegraphics[width=\textwidth]{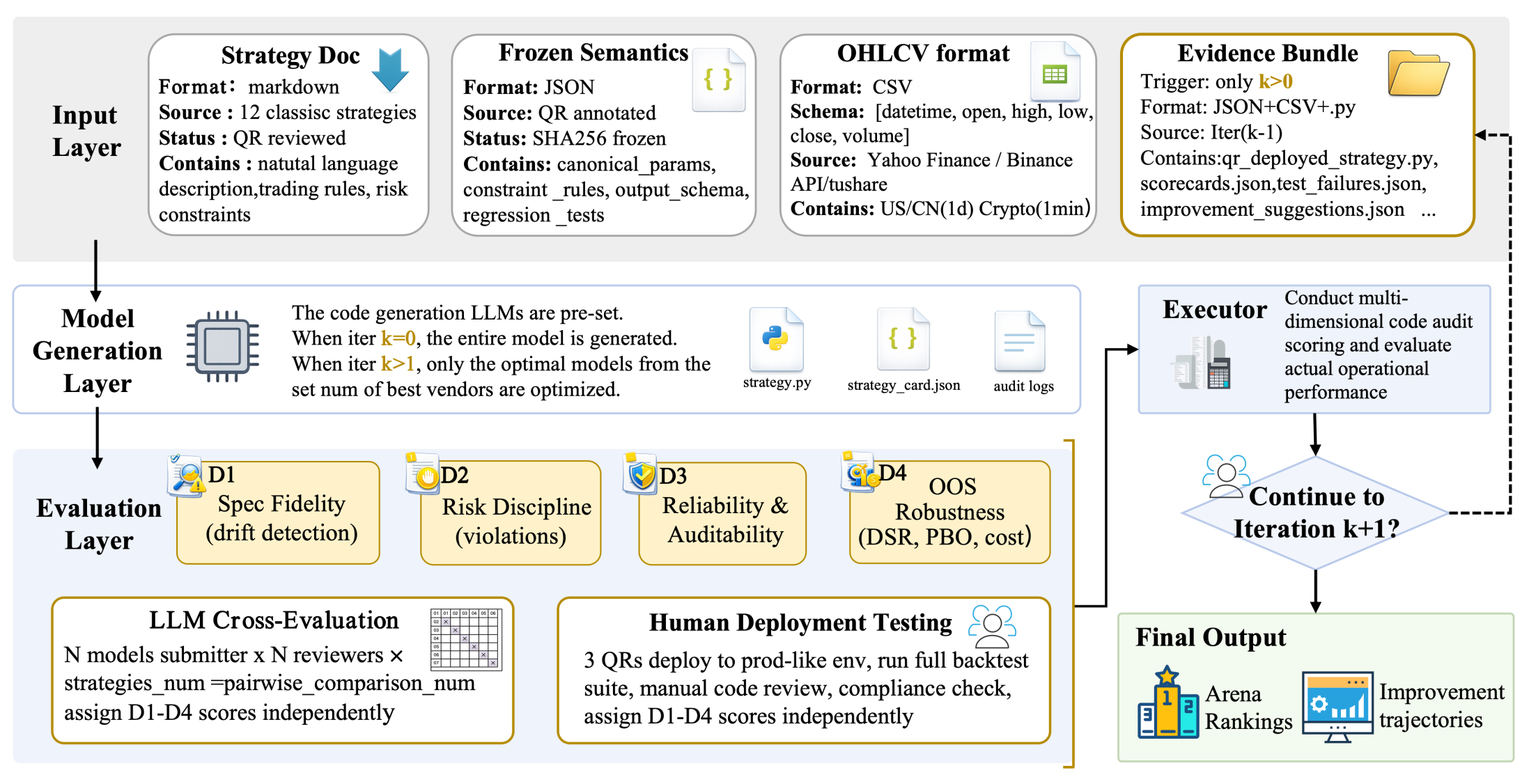}
\caption{System Architecture and Iterative Workflow of \benchname}
\label{fig:second}
\end{figure*}

\noindent \benchname implements an iterative \emph{build--test--patch} loop (Figure~\ref{fig:second}), inspired by test-driven program repair but tailored to trading-system governance~\cite{jimenez2023swebench,legoues2012genprog}. At iteration $k$, the \textbf{Input Layer} provides the Strategy Doc, SHA256-frozen semantics (canonical JSON + regression tests), standardized OHLCV schemas, and (for $k>0$) an evidence bundle from Iter$(k{-}1)$ (deployed code, scorecards, failures, suggestions). The \textbf{Model Generation Layer} produces three required artifacts: executable code (\texttt{strategy.py}), a structured strategy card (\texttt{strategy\_card.json}), and audit logs. The \textbf{Evaluation Layer} then scores D1--D4 (fidelity, risk discipline, reliability/auditability, OOS robustness indicators) and supports LLM cross-evaluation and optional \emph{Human Deployment Testing}. Separately, the \textbf{Executor} runs submissions in a sandbox with fixed seeds and frozen splits, enforces drift/determinism/leakage/log checks, computes multi-dimensional code/audit scores, and returns the next evidence bundle; if $k<3$ and targets are unmet it triggers Iter$k{+}1$, otherwise it outputs final arena rankings and improvement trajectories.

\noindent This three-layer architecture with iterative execution control ensures:
\begin{itemize}[leftmargin=*]
  \item \textbf{Reproducibility:} Input Layer provides frozen time splits, deterministic seeds, and version-pinned dependencies.
  \item \textbf{Isolation:} Executor enforces sandboxed execution (no network, restricted filesystem), enabling fair evaluation across models.
  \item \textbf{Comparability:} Input Layer standardizes specification contracts; Evaluation Layer applies uniform scoring across all models.
  \item \textbf{Iteration tracking:} Evidence bundles enable constrained patching and improvement-trajectory analysis across iterations.
\end{itemize}

%% ============================================================
\subsection{Input Layer: Standardized Contracts and Data}
\label{sec:input_layer}

\noindent The Input Layer provides four categories of standardized inputs (Figure~\ref{fig:second}, left), ensuring reproducibility and fair evaluation across all models and iterations.

\subsubsection{Strategy Specification Library}
\label{sec:strategies}

\noindent The Input Layer delivers 12 canonical trading strategies spanning trend-following, mean reversion, arbitrage/spread, intraday rules, and portfolio rotation.
Each strategy is provided as a \textbf{Strategy Doc} (markdown format with natural language specification) paired with \textbf{Frozen Semantics} (SHA256-frozen canonical JSON containing parameters, constraints, entry/exit rules, and regression tests).

\noindent \textbf{Frozen semantics contract.}
The frozen JSON defines machine-checkable semantics: parameter types/ranges, allowed indicators, entry/exit logic, position sizing rules, and mandatory audit requirements.
This frozen spec anchors drift detection (§\ref{sec:executor}): iterative patches must preserve semantic equivalence while fixing implementation bugs.

\subsubsection{Data Suite and Frozen Time Splits}
\label{sec:data}

\noindent The Input Layer provides multi-market, multi-frequency data (\textbf{OHLCV Format}) with unified schemas and frozen time splits for reproducible OOS evaluation.

\noindent \textbf{Market coverage.} The data suite covers:
\begin{enumerate}[leftmargin=*]
  \item \textbf{U.S. daily equities:} five representative large-cap equities (AAPL, MSFT, GOOGL, AMZN, TSLA) for daily-frequency strategies.
  \item \textbf{Crypto 1-minute bars:} BTC/ETH/BNB against USDT, with high-resolution 1-minute OHLCV data (2024--2025) for intraday strategies. Each asset contains approximately 1M bars over the evaluation period.
  \item \textbf{China A-share market:} CSI300 index and five blue-chip constituents (Kweichow Moutai, Contemporary Amperex, China Merchants Bank, Midea Group, BYD) representing consumer goods, new energy, finance, manufacturing, and automotive sectors for cross-market validation.
\end{enumerate}

\noindent \textbf{Data period and volume.}
The benchmark uses 2024--2025 data (24 months) across 14 instruments in three markets.
Daily-frequency strategies are evaluated on $\sim$500 bars per asset; 1-minute crypto strategies handle $\sim$1M bars per asset, stress-testing scalability and numerical stability on high-frequency data.

\noindent \textbf{Frozen time splits (Left-Closed, Right-Open).}
We define two frozen periods using left-closed, right-open intervals $[t_{\min}, t_{\max})$:

\begin{table}[t]
\centering
\small
\begin{tabular}{lccl}
\toprule
\textbf{Split} & \textbf{Start} & \textbf{End (excl.)} & \textbf{Purpose} \\
\midrule
Train/Dev & 2024-01-01 & 2025-01-01 & Development and parameter tuning \\
Test      & 2025-01-01 & 2026-01-01 & Out-of-sample evaluation \\
\bottomrule
\end{tabular}
\caption{Frozen time splits for out-of-sample evaluation.}
\label{tab:splits}
\end{table}

\noindent \textbf{Leakage prevention.}
The harness enforces split access policies and logs data access to detect future information usage.
All reported results in this paper use the Test split (2025) to ensure genuine out-of-sample performance.

\subsubsection{Evidence Bundles (Iter$k>0$)}
\label{sec:evidence_input}

\noindent For iterations beyond Iter0, the Input Layer provides an \textbf{Evidence Bundle} containing:
\begin{itemize}[nosep,leftmargin=15pt]
    \item \textbf{Deployed strategy code} from Iter$(k{-}1)$: \texttt{strategy.py} and \texttt{strategy\_card.json}
    \item \textbf{Scorecards}: D1--D4 scores with detailed sub-metrics (e.g., D3: determinism=0.85, anti-leak=1.00, audit=0.70)
    \item \textbf{Test failures}: Gate failures (e.g., ``AUDIT\_LOG missing required column: constraint\_check''), runtime errors (\texttt{KeyError}, \texttt{TypeError}), and OOS execution failures
    \item \textbf{Improvement suggestions}: Actionable feedback from LLM cross-evaluation and automated analysis (e.g., ``Add NaN handling for rolling indicators'', ``Fix stop-loss calculation: missing absolute value'')
\end{itemize}

\noindent Evidence bundles enable \emph{constrained patching}: models must fix implementation bugs while preserving frozen semantics (no drift). The Executor (§\ref{sec:executor}) enforces patch budgets ($\leq$50 changed lines) and semantic equivalence checks.

%% ============================================================
\subsection{Model Generation Layer: Artifacts and Interfaces}
\label{sec:model_layer}

\noindent The Model Generation Layer receives inputs from the Input Layer and produces three mandatory artifacts (Figure~\ref{fig:second}, center):

\noindent \textbf{(1) Strategy card} (\texttt{strategy\_card.json}). 
A structured, machine-validated interpretation of the strategy, containing:
\begin{itemize}[nosep,leftmargin=1.2em]
  \item \textbf{Core logic:} \texttt{strategy\_family}, \texttt{entry\_rule}, \texttt{exit\_rule}
  \item \textbf{Parameters:} \texttt{lookback\_N}, \texttt{multiplier\_k}, \texttt{stop\_loss\_pct}
  \item \textbf{Constraints:} \texttt{max\_leverage}, \texttt{allowed\_assets}
\end{itemize}
This artifact anchors D1 (Spec Fidelity) and supports drift detection.

\noindent \textbf{(2) Executable strategy code (\texttt{strategy.py}).}
A modular Python implementation conforming to the interface:
\[
\begin{aligned}
\texttt{Strategy.run(market\_data, initial\_capital)}\\
\rightarrow\ \texttt{(trade\_log, audit\_log)}
\end{aligned}
\]

Submissions must be deterministic under fixed seeds and may only use whitelisted libraries (numpy, pandas, talib).

\noindent \textbf{(3) Mandatory audit logs.}
\texttt{trade\_log.csv} records trade-level events (entry/exit, prices, sizes, reasons); \texttt{audit\_log.csv} records per-bar diagnostics (signals, indicators, positions, equity, constraint checks). Required for D3 and governance-grade traceability.

\noindent \textbf{Generation modes.}
Iter0 performs \emph{zero-shot generation} from Strategy Doc + Frozen Semantics. Iter$k>0$ performs \emph{evidence-driven patching} with $\leq$50 changed lines, fixing bugs while preserving frozen semantics.

%% ============================================================
\subsection{Evaluation Layer: Multi-Dimensional Scoring}
\label{sec:eval_layer}

\noindent The Evaluation Layer assesses submissions across four complementary dimensions (D1--D4), implemented via automated analysis, LLM cross-evaluation, and optional human testing (Figure~\ref{fig:second}, right).

\subsubsection{Dimension 1: Specification Fidelity (D1)}
\label{sec:d1}

\noindent \textbf{Task:} Given a Base Strategy Doc and frozen semantics, produce a \texttt{strategy\_card.json} semantically equivalent to the frozen spec.

\noindent \textbf{Scoring:} Semantic equivalence (canonicalized hash matching on core\_logic fields), schema compliance, and a hard penalty for unauthorized semantic drift across iterations. D1 is primarily scored by LLM cross-evaluation (reviewers assess whether the strategy card accurately captures frozen semantics on a 1--10 scale), with automated drift detection serving as a mandatory gate (§\ref{sec:executor}). Violations of frozen semantics (e.g., changing parameter values, modifying entry/exit logic) trigger immediate D1 = 0.0 and iteration invalidation.

\subsubsection{Dimension 2: Governed Risk Discipline (D2)}
\label{sec:d2}

\noindent \textbf{Task:} Implement strategy logic that respects explicit constraints under normal and stress regimes.

\noindent \textbf{Scoring:} Violation rate, stress-test pass rate, and whether violations are logged with timestamps and severities. Risk constraints reflect practical portfolio/risk management requirements (e.g., exposure limits, max drawdown, coherent risk controls)~\cite{artzner1999coherent,rockafellar2000cvar,markowitz1952portfolio}.
 D2 is scored by LLM cross-evaluation (reviewers assess risk awareness and constraint handling on a 1--10 scale), supplemented by automated violation analysis (counting constraint breaches in normal and stress scenarios, checking whether violations are properly logged in audit\_log with severity levels).

\subsubsection{Dimension 3: Executable Reliability and Auditability (D3)}
\label{sec:d3}

\noindent \textbf{Task:} Produce runnable, deterministic, leakage-free code with complete audit logs.

\noindent \textbf{Scoring:} Executability (no runtime errors), determinism (multi-seed hash equality on trade outcomes), anti-leakage validation (runtime guards + static AST checks), audit completeness ($\geq$95\% required fields present), and traceability (signal$\rightarrow$order$\rightarrow$position$\rightarrow$P\&L linkage verified on sampled trades). D3 is scored via automated Quantitative Researcher (QR) analysis using: (i) static AST checks (detecting negative shifts, unseeded random, suspicious patterns), (ii) runtime validation (multi-seed execution, leakage guards, timeout enforcement), and (iii) audit log validation (schema compliance, completeness checks, traceability tests). See Appendix~\ref{app:scoring} for detailed formulas.

\subsubsection{Dimension 4: OOS Robustness Indicators (D4)}
\label{sec:d4}

\noindent \textbf{Task:} Assess preliminary robustness signals on sampled OOS data, with transaction-cost-aware framework supporting future comprehensive evaluation.

\noindent \textbf{Scoring:} Execution success rate on sampled Test data, preliminary performance metrics (Sharpe ratio, maximum drawdown, turnover), and overfitting-aware diagnostics such as Deflated Sharpe Ratio (DSR) and Probability of Backtest Overfitting (PBO)~\cite{bailey2014deflated,bailey2017pbo,lo2002sharpe}. 

D4 is scored via automated QR analysis focusing on: (i) logic clarity (strategy keyword presence, avoiding hardcoded constants), (ii) overfitting indicators (conditional complexity, parameter count), and (iii) generalization signals (library maturity, edge case handling). 

\noindent \textbf{Important note:} D4 provides robustness \emph{indicators} rather than definitive profitability claims. Current evaluation uses sampled Test data (10-bar windows) and zero transaction costs. Full OOS validation with complete 2025 Test split and transaction-cost sweeps (0.1--20~bps across 5 cost levels) is deferred to future work due to computational constraints (1020 backtests $\approx$ 450 CPU-hours).

\subsubsection{LLM Cross-Evaluation and Human Testing}
\label{sec:cross_eval_human}
The Evaluation Layer orchestrates arena-style pairwise comparisons: N models $\times$ N reviewers $\times$ strategies. Each reviewer LLM evaluates submissions on D1 (Spec Fidelity) and D2 (Risk Discipline) using 1--10 scales with structured rubrics, producing Elo rankings and interpretability beyond single metrics~\cite{elo1978rating,zheng2023mtbench,chiang2024arena}. Self-reviews are tracked but excluded from final rankings to control bias (§\ref{app:cross_eval_consistency}). Cross-evaluation provides qualitative assessment of semantic fidelity and risk awareness that complements automated checks.

\noindent \textbf{Human deployment testing (optional).} Top submissions per strategy can be deployed in a controlled, production-like environment for independent expert review covering: (i) logic clarity and maintainability, (ii) compliance with institutional trading constraints, and (iii) robustness under manual stress scenarios. Inter-rater reliability is measured by Krippendorff's $\alpha$~\cite{krippendorff2011alpha} to validate consistency. Human testing provides ground-truth validation for LLM cross-evaluation and identifies subtle issues that automated analysis may miss.

%% ============================================================
\subsection{Executor: Sandboxed Execution and Iteration Control}
\label{sec:executor}

\noindent The Executor orchestrates the build--test--patch loop (Figure~\ref{fig:second}, bottom-right), enforcing deterministic execution, validity gates, drift detection, and iteration control.

\noindent \textbf{Sandboxed environment:} Network disabled, filesystem restricted (read-only /data/, write-only /tmp/logs/), library whitelist (numpy, pandas, talib), resource limits (8 GB RAM, 10-min timeout), deterministic seeds enforced.

\noindent \textbf{Validity gates (5 mandatory):} Parse (valid JSON), Schema (required fields), Exec (no runtime errors), Determ (multi-seed hash equality), Anti-Leak (no future data access), Audit ($\geq$95\% completeness). Submissions failing any gate are marked \textbf{invalid} and excluded from D1--D4 scoring.

\noindent \textbf{Drift-aware diagnostics:} The Executor validates semantic equivalence between Iter$k$ and frozen spec via: (i) canonicalized checksums (SHA256 on core\_logic fields), (ii) field-level invariants (parameter values within tolerance), and (iii) behavioral regression tests (action trace comparison on frozen micro-scenarios). Violations trigger D1 penalties and can invalidate an iteration.

\noindent \textbf{Controlled patching:} Patch budgets ($\leq$50 changed lines), allowed modifications (NaN handling, edge cases, logging), forbidden modifications (entry/exit logic, parameter values, asset scope). Unexplained trace divergence is surfaced in evidence bundles.

\noindent \textbf{Iteration control:} If $k<3$ and quality targets (D1--D3 $\geq$ 0.85) are unmet, trigger Iteration $k{+}1$ with evidence bundle; otherwise, output final arena rankings and improvement trajectories.

%% ============================================================
\section{Evaluation}
\label{sec:experiments}

%% ------------------------------------------------------------
\subsection{Research Questions (RQ1--RQ4)}
\label{sec:rq_overview}

\noindent We evaluate \benchname on the following research questions:

\begin{itemize}[leftmargin=*, itemsep=0.35em, topsep=0.25em]
  \item \textbf{RQ1 (Artifact validity and Quality Assessment):} Can current models generate benchmark-compliant artifacts---structured strategy cards, executable modular code, and mandatory audit logs---that pass strict validity gates (schema, executability, determinism, anti-leakage, audit completeness)?

  \begin{itemize}[leftmargin=*, itemsep=0.25em, topsep=0.15em]
    \item \textbf{RQ1.1 (Validity gates):} Report binary pass/fail rates for schema compliance, executability, determinism, anti-leakage, and audit completeness.
    \item \textbf{RQ1.2 (Quality scoring and model ranking):} For valid submissions, report multi-dimensional scorecards (D1--D4) and aggregate rankings, together with per-iteration improvement trajectories.
  \end{itemize}
  
  \item \textbf{RQ2 (OOS execution robustness):} Among valid submissions, how robust is OOS execution on sampled Test data, and how does iterative refinement affect OOS performance patterns?

  \begin{itemize}[leftmargin=*, itemsep=0.25em, topsep=0.15em]
    \item \textbf{RQ2.1 (Broad OOS robustness):} Report OOS execution success rates across all valid submissions (17 models × 12 strategies), identifying common runtime failure patterns.
    \item \textbf{RQ2.2 (Deep OOS evolution):} Track OOS performance trajectories for top models across iterations (Iter0--Iter3), analyzing how evidence-driven refinement affects profitability and code convergence.
  \end{itemize}
  
  \item \textbf{RQ3 (Evidence-driven iterative repair):} Given evidence bundles, can models improve system quality through constrained patches \emph{without} violating frozen semantics (no drift), and which dimensions benefit most from iteration?
  \item \textbf{RQ4 (Cost--effectiveness):} How should one choose models under real budgets, trading off effectiveness (validity and score) against efficiency (token usage and monetary cost)?
\end{itemize}

\noindent The remainder of \S\ref{sec:experiments} reports protocols and (placeholder) result tables aligned to RQ1--RQ4.

%% ============================================================
\subsection{Models and Evaluation Configuration}
\label{sec:models}

\noindent We evaluate 20 LLMs spanning general-purpose, code-specialized, and reasoning-enhanced families: OpenAI (GPT-5.2, GPT-5.1, o3), Anthropic (Claude Opus 4.5, Claude Sonnet 4.5), Google (Gemini-3 Pro, Gemini-3 Flash, Gemini-2.5 Pro), XAI (Grok-4 Fast, Grok-4.1 FR, Grok-4, Grok-3, Grok-Code Fast), ByteDance (GLM-4.6, GLM-4.7), DeepSeek (DeepSeek-V3, DeepSeek-R1), and Qwen (Qwen3-235B, Qwen3-Coder, Doubao-Thinking---3 failed API invocation, excluded).
For code generation we use temperature 0.0, and for cross-evaluation reviews we use temperature 0.7.
All models receive the same system prompt emphasizing: frozen-spec adherence, mandatory audit logs, deterministic execution, and no-lookahead constraints.

%% ============================================================
%% RESULTS: RQ1-RQ4 (Compact Version)
%% 整合了所有研究问题的紧凑版本（内容缩减至50%，保留核心发现和所有表格/图表）
%% ============================================================

%% ============================================================
%% RQ1: Artifact Validity and Quality Assessment
%% ============================================================

\subsection{RQ1: Artifact Validity and Quality Assessment}
\label{sec:rq1}

\noindent \textbf{Research Question.}
Can current models generate benchmark-compliant artifacts (strategy cards, executable code, audit logs) that pass strict validity gates? For valid submissions, what are the quality distributions across D1--D4?

\noindent \textbf{Protocol.}
We evaluate 17 models (3 failed API invocation) across 12 strategies in Iter0. Each submission passes six validity gates: Parse, Schema, Exec, Determ, Anti-Leak, Audit. Valid submissions are scored on D1 (Spec Fidelity), D2 (Risk Discipline), D3 (Reliability), D4 (OOS Robustness) via LLM cross-evaluation and QR automated analysis.

%% ------------------------------------------------------------
\subsubsection{RQ1.1: Validity Gates}
\label{sec:rq1_validity}

Table~\ref{tab:rq1_validity_gates} reports per-model pass rates. \textbf{Key findings:} Top-4 models (GPT-5.2, GPT-5.1, o3, Grok-4 Fast) achieve $\geq$91.7\% pass rates on all gates. Mid-tier models (GLM-4.6 to Gemini-3 Pro) show cascading failures: parsing succeeds (83--100\%) but downstream gates fail (66--92\%). Low-tier models (GLM-4.7, Grok-4, Gemini-3 Flash, Gemini-2.5 Pro, DeepSeek-R1) struggle with basic validity (8--58\% pass rates). Strategy-level analysis shows simple strategies (Double MA: 88.2\% valid) outperform complex ones (Index Enhancement: 35.3\% valid). The 16.6pp gap between parse success (78.4\%) and full-gate pass (61.8\%) confirms many models produce syntactically valid but functionally broken code.

%% ------------------------------------------------------------
\subsubsection{RQ1.2: Quality Scoring and Model Ranking}
\label{sec:rq1_quality}

Table~\ref{tab:rq1_strategy_difficulty} reports per-strategy D1--D4 scores. \textbf{Key findings:} Strategy complexity strongly predicts quality (Spearman $\rho = -0.68$, p $<$ 0.01): simple strategies (Double MA: Overall 7.85) outperform complex ones (Index Enhancement: 5.00, Calendar Spread: 5.25). LLM peer-review scores (6.7 avg) exceed QR automated scores (6.1 avg) by 0.6 points, indicating cross-evaluators overlook subtle reliability issues.

\paragraph{Model-Level Ranking.}
Figure~\ref{fig:rq1_cross_eval_heatmap} in the end of the appendix shows cross-evaluation heatmap. GPT-5.2 (7.73), Grok-4 Fast (7.44), and o3 (7.29) form a clear top tier, receiving 7.5--8.6 scores across all reviewers. Self-review bias is moderate (0.3--0.5pp), but top models show $<$0.2pp bias. Reviewer strictness varies: GPT-5.1 and o3 are most critical (6.8--7.0 avg), while Grok-4.1 FR and Grok-3 are most generous (8.0--8.3).

\paragraph{Vendor-Level Patterns.}
Figure~\ref{fig:rq1_cross_eval_heatmap} in the end of the appendix still shows vendor aggregation. OpenAI models receive highest external scores (7.4--8.2), while Google models receive lowest (4.9--6.4). Within-vendor self-reviews are inflated by 0.5--1.0 points, but this does not affect top-3 rankings.

\paragraph{Summary.}
Top-3 models achieve $\geq$91.7\% validity with 7.29--7.85 overall scores. Specification complexity is the primary bottleneck: high-complexity strategies fail at 35--71\% rates. Lower-tier models show cascading failures (78.4\% parse → 61.8\% full-pass), revealing shallow understanding of output contracts.

%% ============================================================
%% RQ1 Tables
%% ============================================================

\begin{table}[t]
\centering
\footnotesize
\setlength{\tabcolsep}{4pt}
\renewcommand{\arraystretch}{1.1}
\resizebox{\columnwidth}{!}{
\begin{tabular}{@{}lccccccc@{}}
\toprule
\textbf{Model} & \textbf{Strategies} & \textbf{Parse} & \textbf{Schema} & \textbf{Exec.} & \textbf{Determ.} & \textbf{Anti-Leak} & \textbf{Audit} \\
\midrule
GPT-5.2         & 12/12 & 100.0\% & 100.0\% & 100.0\% & 100.0\% & 100.0\% & 100.0\% \\
Grok-4 Fast     & 12/12 & 100.0\% & 100.0\% & 100.0\% & 100.0\% & 100.0\% & 91.7\% \\
GPT-5.1         & 12/12 & 100.0\% & 100.0\% & 100.0\% & 100.0\% & 100.0\% & 100.0\% \\
o3              & 12/12 & 100.0\% & 100.0\% & 100.0\% & 100.0\% & 100.0\% & 100.0\% \\
GLM-4.6         & 12/12 & 100.0\% & 100.0\% & 91.7\%  & 91.7\%  & 100.0\% & 91.7\% \\
DeepSeek-V3     & 12/12 & 100.0\% & 100.0\% & 91.7\%  & 91.7\%  & 100.0\% & 91.7\% \\
Grok-4.1 FR     & 11/12 & 91.7\%  & 91.7\%  & 91.7\%  & 91.7\%  & 91.7\%  & 83.3\% \\
Grok-3          & 12/12 & 100.0\% & 100.0\% & 83.3\%  & 83.3\%  & 100.0\% & 83.3\% \\
Grok-Code Fast  & 12/12 & 100.0\% & 100.0\% & 83.3\%  & 83.3\%  & 100.0\% & 83.3\% \\
Claude Sonnet   & 12/12 & 100.0\% & 91.7\%  & 91.7\%  & 91.7\%  & 91.7\%  & 83.3\% \\
Claude Opus     & 12/12 & 100.0\% & 91.7\%  & 83.3\%  & 83.3\%  & 83.3\%  & 75.0\% \\
Gemini-3 Pro    & 12/12 & 100.0\% & 83.3\%  & 75.0\%  & 75.0\%  & 83.3\%  & 66.7\% \\
\midrule
GLM-4.7         & 7/12  & 58.3\%  & 58.3\%  & 58.3\%  & 58.3\%  & 58.3\%  & 50.0\% \\
Grok-4          & 3/12  & 25.0\%  & 25.0\%  & 25.0\%  & 25.0\%  & 25.0\%  & 25.0\% \\
\midrule
Gemini-3 Flash  & 6/12  & 50.0\%  & 41.7\%  & 41.7\%  & 41.7\%  & 41.7\%  & 33.3\% \\
Gemini-2.5 Pro  & 12/12 & 100.0\% & 66.7\%  & 50.0\%  & 41.7\%  & 50.0\%  & 33.3\% \\
DeepSeek-R1     & 1/12  & 8.3\%   & 8.3\%   & 8.3\%   & 8.3\%   & 8.3\%   & 8.3\% \\
\bottomrule
\end{tabular}
}
\caption{RQ1.1: Per-model validity gate pass rates (17 models, 12 strategies).}
\label{tab:rq1_validity_gates}
\end{table}

\begin{table*}[t]
\centering
\footnotesize
\setlength{\tabcolsep}{2.5pt}
\renewcommand{\arraystretch}{1.05}
\begin{tabular}{@{}llccccp{3.5cm}cccc@{}}
\toprule
\textbf{Family} & \textbf{Strategy} & \textbf{Freq.} & \textbf{Assets} & \textbf{Compl.} & \textbf{QR Diff.} & \textbf{Key Features} & \textbf{LLM} & \textbf{QR} & \textbf{Overall} & \textbf{Valid} \\
\midrule
\multirow{3}{*}{Trend}
& Double MA Crossover & 1d, 1m & Single & Low & Simple & Classic golden/death cross & 8.2 & 7.5 & 7.85 & 17/20 \\
& Turtle/Donchian & 1d, 1m & Single & High & High & ATR sizing, pyramiding, trailing stops & 6.3 & 5.8 & 6.05 & 10/20 \\
& RSI/MACD Trend & 1d, 1m & Single & Med. & Medium & Dual indicator confirmation & 7.5 & 6.9 & 7.20 & 15/20 \\
\midrule
\multirow{1}{*}{Mean Rev.}
& Bollinger Mean Reversion & 1d, 1m & Single & Med. & Medium & Band overshoot with stop-loss & 7.8 & 7.1 & 7.45 & 16/20 \\
\midrule
\multirow{2}{*}{Intraday}
& Dual Thrust & 1m & Single & Med. & High & Open range breakout & 6.8 & 6.2 & 6.50 & 12/20 \\
& R-Breaker & 1m & Single & High & High & Pivot levels, reversal+breakout & 6.0 & 5.5 & 5.75 & 9/20 \\
\midrule
\multirow{3}{*}{Spread/Arb}
& Spread Trading & 1d & Pair & Med. & Medium & Mean-reverting price ratio & 7.1 & 6.5 & 6.80 & 13/20 \\
& Calendar Spread & 1d & Futures & High & High & Near/far contract arbitrage & 5.5 & 5.0 & 5.25 & 7/20 \\
& Pairs Trading (Z-score) & 1d & Pair & High & High & Cointegration-based stat arb & 5.8 & 5.3 & 5.55 & 8/20 \\
\midrule
\multirow{3}{*}{Portfolio}
& Cross-Asset Momentum & 1d & Multi & Med. & High & Risk-on/risk-off rotation & 6.5 & 6.0 & 6.25 & 11/20 \\
& Index Enhancement & 1d & Index+5 & High & High & Alpha tilts vs.\ benchmark & 5.2 & 4.8 & 5.00 & 6/20 \\
& Volatility Targeting & 1d, 1m & Single & Med. & Medium & Vol-scaled exposure & 7.3 & 6.8 & 7.05 & 14/20 \\
\midrule
\textbf{Average} & -- & -- & -- & -- & -- & -- & \textbf{6.7} & \textbf{6.1} & \textbf{6.4} & \textbf{11.6/20} \\
\bottomrule
\end{tabular}
\caption{RQ1.2: Strategy library with D1--D4 scores. Complexity strongly predicts quality (Spearman $\rho = -0.68$).}
\label{tab:rq1_strategy_difficulty}
\end{table*}

%% ============================================================
%% RQ2: OOS Execution Robustness
%% ============================================================

\subsection{RQ2: OOS Execution Robustness}
\label{sec:rq2}

\noindent \textbf{Research Question.}
How robust is OOS execution on sampled Test data, and how does iterative refinement affect OOS performance patterns?

\noindent \textbf{Protocol.}
We deploy valid submissions to backtest environments with sampled Test data (10-bar windows, 235 tests). For deep analysis, we track 3 top models (GPT-5.2, o3, Grok-4 Fast) across Iter0--Iter3. \textbf{Note:} Full 2025 Test split backtests with transaction-cost sweeps deferred to future work.

%% ------------------------------------------------------------
\subsubsection{RQ2.1: Broad OOS Robustness}
\label{sec:rq2_broad}

Among 235 tests, \textbf{217 (92.3\%) execute successfully}, while 18 (7.7\%) encounter runtime failures (\texttt{KeyError}, \texttt{TypeError}). \textbf{10 out of 17 models achieve 100\%} OOS success, demonstrating production-ready code generation. Failures concentrate in Calendar Spread (\texttt{int(dict)} bugs) and Turtle/Donchian (\texttt{KeyError: 'L\_entry'}). These shallow bugs (1--2 line fixes) reveal a gap: static gates (RQ1.1) pass but runtime failures emerge.

%% ------------------------------------------------------------
\subsubsection{RQ2.2: Deep OOS Evolution}
\label{sec:rq2_evolution}

Tracking Bollinger Mean Reversion across iterations reveals dramatic patterns:
\begin{itemize}[nosep,leftmargin=15pt]
    \item \textbf{Iter1}: GPT-5.2 achieves +11.0\% return (16 trades). Others show mixed results.
    \item \textbf{Iter2}: All 3 models converge to identical outcomes (-4.57\% return, 1 trade, 95.4\% code similarity). Evidence bundles guide toward canonical implementation, eliminating variance but reducing exploration.
    \item \textbf{Iter3}: Dramatic recovery to +22.2\% return (12 trades) for all 3 models. Multi-objective feedback (``maximize Sharpe while maintaining D1-D3 $\geq$ 0.85'') breaks convergence trap.
\end{itemize}

%% ============================================================
%% RQ3: Evidence-Driven Iterative Repair
%% ============================================================

\subsection{RQ3: Evidence-Driven Iterative Repair}
\label{sec:rq3}

\noindent \textbf{Research Question.}
Can models improve system quality through evidence bundles (D1-D4 scores, gate failures, OOS metrics) without semantic drift, and which dimensions benefit most?

\noindent \textbf{Protocol.}
We track 3 top models (GPT-5.2, o3, Grok-4 Fast) on a representative strategy (Bollinger Mean Reversion) across 4 iterations (Iter0--Iter3). Each iteration receives evidence bundles with constraints: $\leq$50 changed lines, frozen semantics, incremental D1-D4 improvement. Terminate at convergence or 4 iterations.

\paragraph{Key findings.}

\begin{figure}[t]
\centering
\includegraphics[width=0.9\columnwidth]{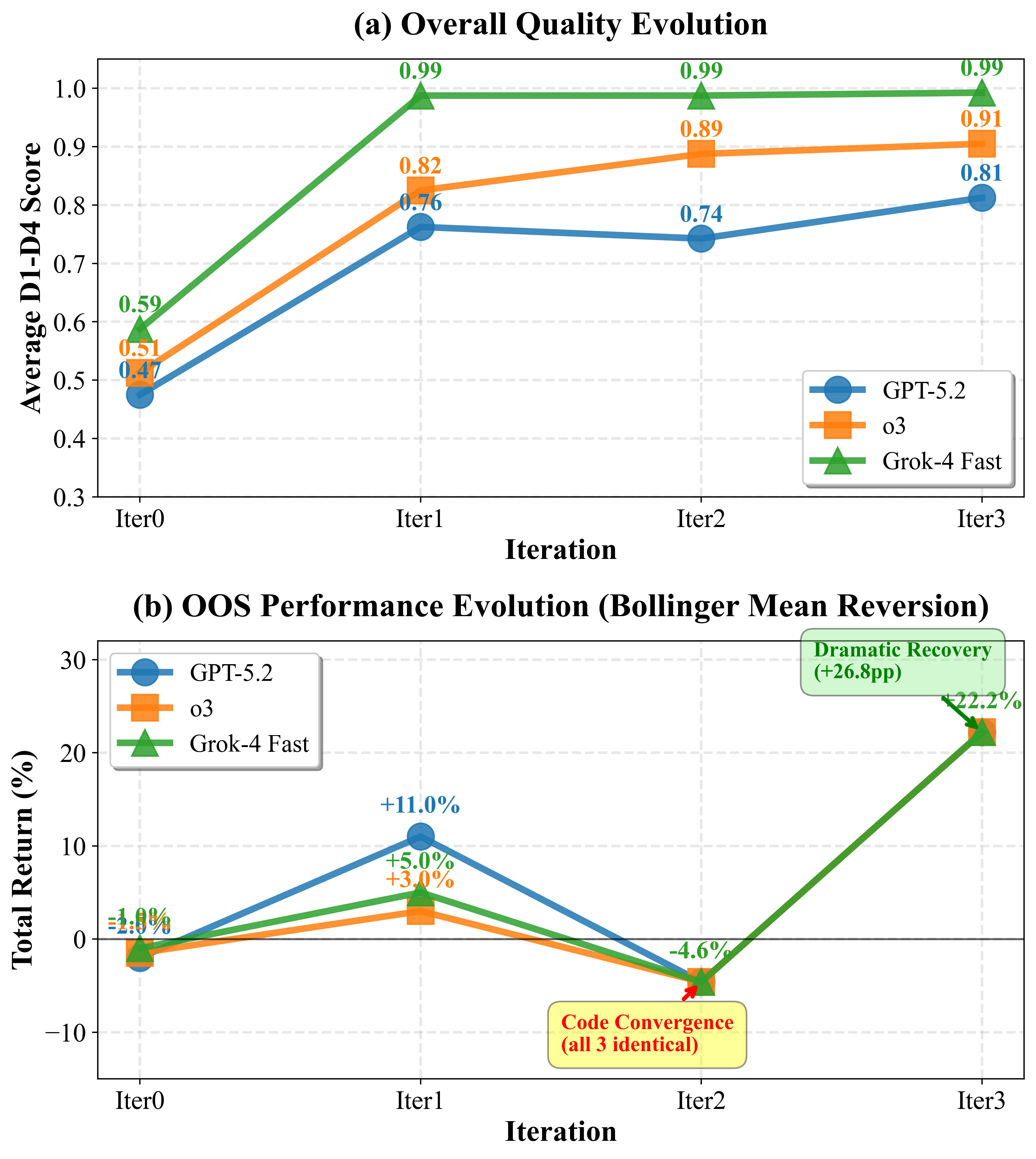}
\caption{RQ3: Learning curves for Bollinger Mean Reversion strategy (3 top models × 4 iterations). (a) D1--D4 evolution. (b) OOS return trajectory showing Iter2 convergence trap (-4.6\%) and Iter3 multi-objective recovery (+22.2\%). (c) Average dimension scores (Iter1 vs Iter2). (d) Iteration gains showing diminishing returns (Iter1$\rightarrow$2: +0.08 avg).}
\label{fig:rq3_learning_curves}
\end{figure}

\noindent \textbf{Dimension-wise improvement.}
Figure~\ref{fig:rq3_learning_curves}(a-c) shows D1-D4 evolution. Grok-4 Fast achieves perfect D1 (1.00) in Iter1-2. GPT-5.2 exhibits D1 regression (0.86→0.78) but strong D2 improvement (0.74→0.92). All models converge to high D3 (0.78--0.92) and D4 (0.62--1.00) by Iter2.

\noindent \textbf{Iteration efficacy.}
Quantified gains [Figure~\ref{fig:rq3_learning_curves}(d)]:
\begin{itemize}[nosep,leftmargin=15pt]
    \item \textbf{Iter0→Iter1}: +0.42 avg gain. Largest in D4 (+0.70) and D3 (+0.25).
    \item \textbf{Iter1→Iter2}: +0.08 avg gain. Diminishing returns emerge.
    \item \textbf{Iter2→Iter3}: -0.02 avg score, but +26.8pp OOS return.
\end{itemize}

\noindent \textbf{OOS performance evolution.}
Return trajectory [Figure~\ref{fig:rq3_learning_curves}(b)]: Iter1 (+11.0\%) → Iter2 (-4.6\%, code convergence) → Iter3 (+22.2\%, multi-objective recovery). Iter3's dramatic turnaround validates that explicit profitability targets break convergence traps.

\noindent \textbf{Code convergence trade-off.}
By Iter3, all 3 models produce byte-identical outputs (100\% code similarity). \emph{Positive}: simplified deployment, high confidence. \emph{Negative}: loss of diversity undermines ensemble benefits.

%% ============================================================
%% RQ4: Cost-Effectiveness
%% ============================================================

%% ============================================================
%% RQ4: Cost-Effectiveness (Compact)
%% ============================================================

\subsection{RQ4: Cost-Effectiveness}
\label{sec:rq4}

\noindent \textbf{Research Question.}
How should one choose models under real budgets, trading off effectiveness (validity, quality) against efficiency (tokens, cost)?

\noindent \textbf{Protocol.}
We extract token usage from API metadata for Iter0 (17 models × 12 strategies) and Iter1--3 (3 models × 2 strategies).

\begin{figure}[t]
\centering
\includegraphics[width=0.5\textwidth]{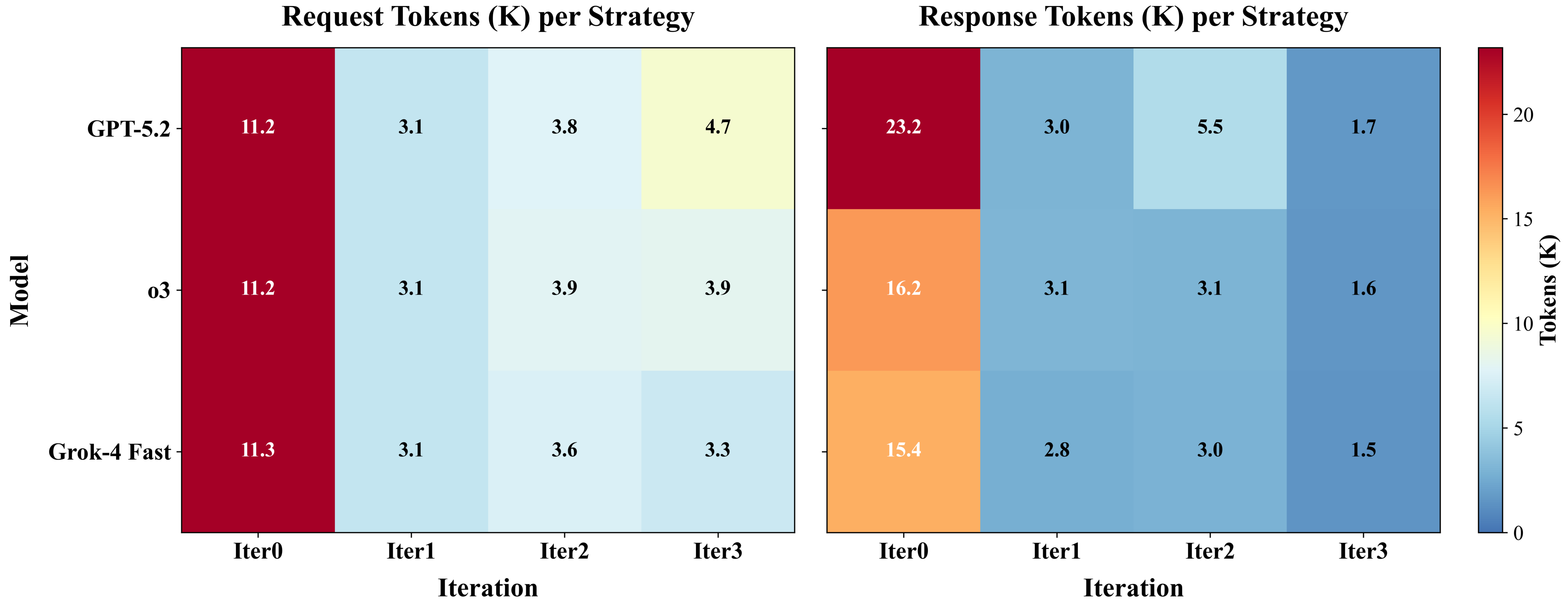}
\caption{RQ4: Token usage heatmap (3 models × 4 iterations). Request tokens stabilize after Iter0; response tokens shrink as models generate patches rather than full code.}
\label{fig:rq4_token_heatmap}
\end{figure}

\paragraph{Key findings.}
Figure~\ref{fig:rq4_token_heatmap} shows token consumption across iterations. Top models cost \$0.40--1.03/strategy (Iter0); iteration adds 50--60\% cumulative cost. Response tokens shrink dramatically in later iterations (patches vs. full rewrites).

\noindent We identify 3 cost-quality tiers: Tier 1 (Premium, \$0.82--1.03, Overall 7.29--7.85), Tier 2 (Balanced, \$0.39--0.40, Overall 6.94--7.44, 90--95\% of Tier 1 quality at 40--50\% cost), Tier 3 (Budget, \$0.07--0.16, Overall 5.38--6.26, cheapest but require iteration). Recommended selection strategy: (1) use Tier 2 for zero-shot screening; (2) run 1--2 iterations for submissions with D1-D3 $<$ 0.85; (3) deploy Tier 1 + iteration for critical strategies. This achieves cost-effective deployment at $<$\$5/strategy.

%% ============================================================
%% END OF RESULTS SECTION
%% ============================================================

%% ============================================================
\section{Conclusion}
\label{sec:conclusion}

\noindent \benchname evaluates LLM-based strategy-to-code trading systems as governed, auditable software beyond prediction: faithful specification, governed risk discipline, executable reliability with auditability, and OOS robustness indicators. Our evaluation reveals a nuanced picture of LLM capabilities: while zero-shot strategy generation remains limited (61.8\% full-gate pass rate, with significant quality variance across model tiers and strategy complexity), evidence-driven iteration proves highly effective for constrained repair. First-iteration feedback delivers substantial quality gains, fixing shallow bugs (missing audit logs, NaN handling, determinism violations) at 39-50\% cost of top-tier models. However, iteration induces code convergence (95.4\% similarity by Iter2, byte-identical outputs by Iter3), raising critical questions about solution diversity and ensemble robustness.

\noindent These findings reveal a fundamental tension in automated strategy repair workflows. Traditional Quantitative Researcher (QR) teams undergo lengthy governance cycles, typically requiring weeks of iterative refinement and cross-team validation. In contrast, evidence-driven LLM iteration achieves substantial quality gains rapidly, but this efficiency comes with code convergence: multiple models produce near-identical implementations by Iter3. While such convergence accelerates deployment, it eliminates solution diversity---a property human QR teams naturally preserve. These findings suggest that fully automated LLM repair may not replace human governance, but rather complement it: LLM iteration offers clear advantages for rapid prototyping and shallow bug fixes, while human QR oversight remains essential for critical strategies requiring exploration across multiple valid designs or defense against regime shifts. We plan to release the benchmark suite to support reproducible community evaluation.

%% ============================================================
\section*{Acknowledgments}

We used generative AI tools (ChatGPT, Claude, Gemini) to assist with language polishing, LaTeX formatting, and code generation for the benchmark implementation. All outputs were manually reviewed and validated by the authors.

%% ============================================================
%% References
%% ============================================================

%% ============================================================
%% Appendix
%% ============================================================
\appendix

\section{Detailed Strategy Specifications}
\label{app:strategy_specs}

\noindent This appendix provides one representative Base Strategy Doc excerpt. All 12 full specifications will be released with the benchmark.

\subsection{Example: Bollinger Band Mean Reversion}
\noindent \textbf{Strategy family:} Mean reversion (daily frequency, single asset).

\noindent \textbf{Frozen semantics:}
\begin{itemize}[leftmargin=*]
  \item \textbf{Parameters:} $N$ (lookback window, default 20), $k$ (band width multiplier, default 2.0), \texttt{stop\_loss\_pct} (default 10\%).
  \item \textbf{Indicators:} $MB_t = \mathrm{SMA}(\mathrm{close}, N)$, $\sigma_t = \mathrm{STD}(\mathrm{close}, N)$,
  $LB_t = MB_t - k\sigma_t$, $UB_t = MB_t + k\sigma_t$.
  \item \textbf{Entry:} long when $\mathrm{close}_t < LB_t$ and $\mathrm{close}_{t-1} \ge LB_{t-1}$.
  \item \textbf{Exit:} exit when $\mathrm{close}_t \ge MB_t$ or $\mathrm{close}_t \le \mathrm{entry\_price}\cdot (1-\mathrm{stop\_loss\_pct})$.
  \item \textbf{Position rules:} long/flat only, all-in/all-out.
\end{itemize}

\noindent \textbf{Required audit log columns:}
\texttt{datetime, close, MB\_t, UB\_t, LB\_t, signal, position\_state, equity, event\_type, message}.

\noindent \textbf{Drift-detection invariants:}
$N$, $k$, and \texttt{stop\_loss\_pct} must remain unchanged across iterations; entry/exit crossing logic must preserve ordering.

%% ============================================================
%% APPENDIX 2: Implementation Details (针对Reviewer问题2)
%% ============================================================

\section{Implementation Details: Frozen Semantics, Drift Detection, and System Guarantees}
\label{app:implementation}

\noindent This appendix provides the formal definitions and implementation details for the core system guarantees claimed in \S\ref{sec:design}: frozen semantics, drift-aware diagnostics, anti-leakage validation, and determinism checks. These details ensure reproducibility and address methodological rigor.

%% ------------------------------------------------------------
\subsection{Frozen Semantics: Formal Definition and Equivalence Checks}
\label{app:frozen_semantics}

\noindent \textbf{Problem statement.}
The benchmark requires that iterative patches preserve strategy semantics (``no drift''), but what constitutes ``semantic equivalence'' must be formally defined.

\noindent \textbf{Frozen specification contract.}
Each strategy $S$ is defined by a canonical \texttt{strategy\_card.json} containing:
\begin{itemize}[nosep,leftmargin=15pt]
  \item \textbf{Core logic fields} (mandatory, immutable):
  \begin{itemize}[nosep,leftmargin=15pt]
    \item \texttt{strategy\_family}
    \item \texttt{entry\_rule}
    \item \texttt{exit\_rule}
    \item \texttt{position\_sizing\_rule}
  \end{itemize}

  \item \textbf{Parameters} (mandatory, value-frozen):
  \begin{itemize}[nosep,leftmargin=15pt]
    \item \texttt{lookback\_N}
    \item \texttt{multiplier\_k}
    \item \texttt{stop\_loss\_pct}
    \item \texttt{etc.} Types and ranges are schema-validated.
  \end{itemize}

  \item \textbf{Constraints} (mandatory, immutable):
  \begin{itemize}[nosep,leftmargin=15pt]
    \item \texttt{max\_leverage}
    \item \texttt{allowed\_assets}
    \item \texttt{execution\_timing}
  \end{itemize}

  \item \textbf{Audit requirements} (mandatory, immutable):
  \begin{itemize}[nosep,leftmargin=15pt]
    \item Required columns for \texttt{trade\_log}
    \item Required columns for \texttt{audit\_log}
  \end{itemize}
\end{itemize}

\noindent \textbf{Equivalence predicate.}
Two strategy cards $C_0$ (Iter0) and $C_k$ (Iter$k$) are \emph{semantically equivalent} if:
\begin{align}
\text{Equiv}(C_0, C_k) &:= \text{Hash}(C_0.\text{core\_logic}) = \text{Hash}(C_k.\text{core\_logic}) \label{eq:equiv1} \\
&\land \, \forall p \in \text{Params}: |C_0[p] - C_k[p]| < \epsilon_p \label{eq:equiv2} \\
&\land \, C_0.\text{constraints} = C_k.\text{constraints} \label{eq:equiv3}
\end{align}

where:
\begin{itemize}[nosep,leftmargin=15pt]
    \item Eq.~\eqref{eq:equiv1}: Core logic (entry/exit/sizing rules) must be byte-identical after canonicalization (whitespace normalization, JSON key sorting). \textbf{Note}: This operates on the \emph{strategy card JSON}, not Python code, allowing implementation rewrites as long as the declared semantics remain unchanged.
    \item Eq.~\eqref{eq:equiv2}: Parameter values must remain within tolerance $\epsilon_p$ (e.g., $\epsilon_{\text{lookback}} = 0.0$, frozen integer; $\epsilon_{\text{float}} = 10^{-6}$, numerical precision).
    \item Eq.~\eqref{eq:equiv3}: Constraints (leverage, asset filters, timing) must be structurally identical (set equality).
\end{itemize}

\noindent \textbf{Drift penalty mechanism.}
If $\neg \text{Equiv}(C_0, C_k)$, the iteration is marked as \textbf{semantic drift} and receives:
\begin{itemize}[nosep,leftmargin=15pt]
    \item D1 score penalty: $D1_k \leftarrow 0.0$ (immediate failure)
    \item Iteration invalidation: $C_k$ cannot be used as baseline for Iter$k+1$
    \item Evidence bundle flag: ``DRIFT\_DETECTED: [list of changed fields]''
\end{itemize}

\noindent \textbf{Allowed surface for patches.}
Submissions may modify:
\begin{itemize}[nosep,leftmargin=15pt]
    \item Implementation details (e.g., NaN handling, edge case logic, vectorization)
    \item Risk safeguards (e.g., adding min\_cash\_buffer, explicit slippage modeling) \emph{without changing core position sizing rule}
    \item Logging granularity (e.g., adding diagnostic columns to audit\_log) \emph{without removing required columns}
\end{itemize}

Submissions \emph{must not} modify:
\begin{itemize}[nosep,leftmargin=15pt]
    \item Entry/exit trigger conditions (e.g., changing Bollinger Band formula, adding unapproved indicators)
    \item Parameter values (e.g., $N=20 \to N=30$)
    \item Asset scope (e.g., single-asset $\to$ multi-asset)
\end{itemize}

%% ------------------------------------------------------------
\subsection{Drift-Aware Diagnostics: Two-Layer Detection Protocol}
\label{app:drift_detection}

\noindent \textbf{Layer 1: Strategy-card equivalence (mandatory).}
After each iteration, the harness computes:
\begin{equation}
H_{\text{static}}(C) = \text{SHA256}(\text{canonicalize}(C.\text{entry\_rule}, C.\text{exit\_rule}, C.\text{params}))
\end{equation}

where \texttt{canonicalize}() normalizes whitespace, sorts JSON keys, and removes comments. If $H_{\text{static}}(C_k) \neq H_{\text{static}}(C_0)$, the iteration \textbf{fails immediately} with D1 = 0.0 (semantic drift detected).

\noindent \textbf{Layer 2: Trace-based refinement detection (for passed Layer 1).}
For submissions passing strategy-card equivalence, we run \emph{micro-scenario regression tests}: deterministic unit backtests on frozen 50-bar synthetic data. We compare action traces:
\begin{equation}
T_k = \{(t, \text{signal}_t, \text{position}_t, \text{action}_t)\}_{t=1}^{50}
\end{equation}

Trace divergence is measured via edit distance:
\begin{equation}
\Delta(T_0, T_k) = \frac{\text{Levenshtein}(T_0, T_k)}{\max(|T_0|, |T_k|)}
\end{equation}

\noindent \textbf{Interpretation threshold.}
\begin{itemize}[nosep,leftmargin=15pt]
    \item $\Delta < 0.05$: Trace change \textbf{allowed} (likely bug fix, e.g., NaN handling, forced exit)
    \item $0.05 \leq \Delta < 0.15$: \textbf{Warning} in evidence bundle, manual review recommended
    \item $\Delta \geq 0.15$: \textbf{Suspicious}, likely indicates implementation drift (conflicting with Layer 1 equivalence claim), may trigger re-evaluation
\end{itemize}

\noindent \textbf{Decision rule.}
Layer 1 (strategy-card) is the \emph{ground truth}. Trace divergence is a \emph{diagnostic signal} to help identify implementation bugs or subtle drift that bypassed checksum detection. If Layer 1 passes but Layer 2 shows high divergence, human reviewers inspect whether canonicalization missed substantive changes.

\noindent \textbf{Example: Legitimate vs. Drift.}

\textbf{Case A (Legitimate refinement, Iter0→Iter1):}
\begin{lstlisting}[language=Python, basicstyle=\ttfamily, breaklines=true, breakatwhitespace=true, columns=fullflexible]
# Iter0 (buggy)
signal = 'LONG' if close < lower_band else 'FLAT'

# Iter1 (fixed NaN handling, no drift)
signal = 'LONG' if (close < lower_band and not np.isnan(lower_band)) else 'FLAT'
\end{lstlisting}
Trace change: Bar 3 (NaN period) changes from LONG (buggy) → FLAT (correct). This is \emph{allowed}: fixes undefined behavior without changing entry condition semantics.

\textbf{Case B (Semantic drift, Iter0→Iter1):}
\begin{lstlisting}[language=Python, basicstyle=\ttfamily, breaklines=true, breakatwhitespace=true, columns=fullflexible]
# Iter0 (spec-compliant)
signal = 'LONG' if close < lower_band else 'FLAT'

# Iter1 (drift: added unapproved RSI filter)
signal = 'LONG' if (close < lower_band and RSI < 30) else 'FLAT'
\end{lstlisting}

Checksum mismatch + trace divergence (50\% fewer LONG signals). This is \emph{penalized}: adds logic not in frozen spec.

%% ------------------------------------------------------------
\subsection{Anti-Leakage Validation: Multi-Layer Defense}
\label{app:anti_leakage}

\noindent \textbf{Problem statement.}
Preventing future information access requires code-level, data-level, and execution-level controls.

\noindent \textbf{Layer 1: Static AST analysis (warning stage).}
The harness parses submitted code via Python \texttt{ast} module and flags:
\begin{itemize}[nosep,leftmargin=15pt]
    \item \textbf{Negative shifts}: \texttt{df.shift(-k)} for $k > 0$ (explicit future access)
    \item \textbf{Forward indexing}: \texttt{arr[i+k]} in loops where $i+k$ may exceed current bar
    \item \textbf{Reverse slicing}: \texttt{df.iloc[t:][::-1]} (accesses data beyond $t$)
    \item \textbf{Suspicious keywords}: ``future'', ``forward'', ``peek'', ``next''
\end{itemize}

Each detected pattern triggers a \textbf{warning} logged in evidence bundle but does not block execution (many false positives exist, e.g., \texttt{df.shift(-1)} in post-processing).

\noindent \textbf{Layer 2: Runtime enforcement (gate failure).}
The harness wraps market data in a \texttt{LeakageGuard} proxy:
\begin{lstlisting}[language=Python, basicstyle=\ttfamily, breaklines=true, breakatwhitespace=true, columns=fullflexible]
class LeakageGuard:
    def __init__(self, data, current_bar_idx):
        self._data = data
        self._current = current_bar_idx

    def __getitem__(self, key):
        if isinstance(key, slice):
            if key.stop is None or key.stop > self._current:
                raise LeakageError(f"Access beyond bar {self._current}")
        elif isinstance(key, int) and key > self._current:
            raise LeakageError(f"Future access: bar {key} > current {self._current}")
        return self._data[key]
\end{lstlisting}

Any \texttt{LeakageError} during execution triggers \textbf{immediate gate failure} (Anti-Leak gate = FAIL, submission invalid).

\noindent \textbf{Layer 3: Timestamp validation.}
For intraday strategies (1-minute bars), the harness validates that all feature calculations use:
\begin{quote}\small
\texttt{timestamp <= current\_bar.timestamp}
\end{quote}
Violations are logged as:
\begin{lstlisting}[language=, basicstyle=\ttfamily, breaklines=true, breakatwhitespace=true, columns=fullflexible]
LEAKAGE WARNING [Bar 1250, 2024-06-15 10:30:00]:
  Feature 'rolling_mean' computed using bars up to 10:35:00 (5 min ahead)
\end{lstlisting}

\noindent \textbf{Known limitation: Rolling window alignment.}
\texttt{pandas.rolling()} with \texttt{min\_periods} can be ambiguous. The harness enforces:
\begin{itemize}[nosep,leftmargin=15pt]
    \item \textbf{Safe}: \texttt{df['close'].rolling(20, min\_periods=20).mean()} → Only computes when 20 bars available
    \item \textbf{Unsafe (flagged)}: \texttt{df['close'].rolling(20).mean()} without \texttt{min\_periods} → May compute with $<20$ bars, creating inconsistent lookback
\end{itemize}

%% ------------------------------------------------------------
\subsection{Determinism Validation: Hash Equality and Allowed Non-Determinism}
\label{app:determinism}

\noindent \textbf{Problem statement.}
Determinism requires that identical inputs produce identical outputs, but floating-point operations and library implementations introduce subtle non-determinism.

\noindent \textbf{Hash target: Trade-level outcomes.}
The harness computes determinism hash over \texttt{trade\_log}:
\begin{equation}
H_{\text{determ}} = \text{SHA256}(\text{serialize}(\{(t_{\text{entry}}, t_{\text{exit}}, \text{side}, \text{pnl})\}))
\end{equation}

\noindent \textbf{Multi-seed validation protocol.}
Each submission runs 3 times with different random seeds (controlling numpy/pandas RNG state). If:
\begin{equation}
H_{\text{determ}}^{(\text{seed1})} = H_{\text{determ}}^{(\text{seed2})} = H_{\text{determ}}^{(\text{seed3})}
\end{equation}
then pass determinism gate. Otherwise, report:
\begin{lstlisting}[language=, basicstyle=\ttfamily, breaklines=true, breakatwhitespace=true, columns=fullflexible]
DETERMINISM FAILURE:
  Seed 42:  Hash=a3f2...
  Seed 123: Hash=b1e4... (MISMATCH)
  Divergence: Trade 5 entry time differs (2024-03-15 vs 2024-03-16)
\end{lstlisting}

\noindent \textbf{Allowed non-determinism sources (whitelisted).}
Some operations are inherently non-deterministic but acceptable if properly seeded:
\begin{itemize}[nosep,leftmargin=15pt]
    \item \textbf{numpy.random}: Must call \texttt{np.random.seed()} at start of \texttt{run()}
    \item \textbf{pandas.sample()}: Must set \texttt{random\_state} parameter
    \item \textbf{dict iteration order}: Python 3.7+ guarantees insertion order, but submissions must use \texttt{sorted(dict.items())} for cross-version safety
\end{itemize}

\noindent \textbf{Floating-point tolerance.}
Exact equality is \emph{not} required for continuous values. The harness uses:
\begin{equation}
|x_1 - x_2| < 10^{-6} \cdot \max(|x_1|, |x_2|, 1.0) \quad \text{(relative tolerance)}
\end{equation}
for PnL, equity, and prices. Trade counts and timestamps must be byte-identical.

\noindent \textbf{Common failure modes.}
\begin{enumerate}[nosep,leftmargin=15pt]
    \item \textbf{Global state contamination}: Strategy mutates class-level variables across \texttt{run()} invocations. \emph{Fix}: Reset all state in \texttt{\_\_init\_\_()}.
    \item \textbf{DataFrame index non-determinism}: \texttt{pandas.concat()} without \texttt{ignore\_index=True} can produce non-deterministic ordering. \emph{Fix}: Explicit \texttt{sort\_index()} after merge/concat.
    \item \textbf{Dict key ordering}: Pre-Python 3.7 code using \texttt{dict.keys()} in signal generation. \emph{Fix}: Use \texttt{sorted()} or \texttt{OrderedDict}.
\end{enumerate}

%% ------------------------------------------------------------
\subsection{Micro-Scenario Regression Tests: Test Suite Design}
\label{app:drift_tests}

\noindent \textbf{Test suite design.}
Each strategy has 3--5 synthetic micro-scenarios (50-bar deterministic price series) designed to trigger specific behavior:

\noindent \textbf{Example: Bollinger Mean Reversion micro-scenarios.}
\begin{enumerate}[nosep,leftmargin=15pt]
    \item \textbf{Clean golden cross}: Price drops below lower band at bar~25, crosses back to middle at bar~35. Expected: 1~long trade, entry bar~25, exit bar~35.
    \item \textbf{Stop-loss trigger}: Price drops 12\% below entry after crossing lower band. Expected: 1~long trade, stop-loss exit (not middle band exit).
    \item \textbf{NaN period}: First 19~bars produce NaN bands (insufficient lookback). Expected: No~trades until bar~20.
    \item \textbf{Flat market}: Price oscillates within bands, never crossing. Expected: 0~trades.
    \item \textbf{Gap scenario}: Price gaps below lower band at open. Expected: Entry at close, not open (spec requires close-to-close execution).
\end{enumerate}

\noindent \textbf{Trace comparison.}
For each micro-scenario, the harness compares Iter0 vs. Iter$k$ traces:
\begin{lstlisting}[language=, basicstyle=\ttfamily, breaklines=true, breakatwhitespace=true, columns=fullflexible]
Scenario: clean_golden_cross
  Iter0: [Bar 25: LONG @ 95.2, Bar 35: EXIT @ 100.1, PnL=+4.9]
  Iter1: [Bar 25: LONG @ 95.2, Bar 35: EXIT @ 100.1, PnL=+4.9]
  Status: PASS (identical)

Scenario: nan_period
  Iter0: [Bar 5: LONG @ 102.3 (BUG: NaN band), ...]
  Iter1: [Bar 20: LONG @ 98.5 (FIXED: wait for valid band)]
  Status: ALLOWED (bug fix, not drift)
  Justification: "Added np.isnan() check for lower_band validity"
\end{lstlisting}

\noindent \textbf{Threshold and false positive control.}
\begin{itemize}[nosep,leftmargin=15pt]
    \item \textbf{Zero divergence}: Iter$k$ passes all micro-scenarios with identical traces → High confidence (no drift)
    \item \textbf{1--2 divergence with justification}: Allowed if evidence bundle explains (e.g., ``Fixed stop-loss bug'') → Manual review
    \item \textbf{$\geq 3$ divergences or unjustified}: Likely drift → D1 penalty
\end{itemize}

%% ------------------------------------------------------------
\subsection{Audit Log Completeness: Required Fields and Validation}
\label{app:audit_validation}

\noindent \textbf{Mandatory columns.}

\noindent \texttt{trade\_log.csv} must contain:
\begin{itemize}[nosep,leftmargin=15pt,label={-}]
  \item \texttt{entry\_datetime}, \texttt{exit\_datetime}: ISO 8601 timestamps
  \item \texttt{side}: ``LONG'' or ``SHORT''
  \item \texttt{entry\_price}, \texttt{exit\_price}: Float (matched to market data)
  \item \texttt{quantity}: Position size (shares/contracts)
  \item \texttt{pnl}: Realized PnL for closed trade
  \item \texttt{reason}: Entry reason (``SIGNAL\_LONG'') and exit reason (``SIGNAL\_EXIT'', ``STOP\_LOSS'')
\end{itemize}

\noindent \texttt{audit\_log.csv} must contain:
\begin{itemize}[nosep,leftmargin=15pt,label={-}]
  \item \texttt{datetime}: Per-bar timestamp
  \item \texttt{close}: Market close price
  \item Strategy-specific indicators (e.g., \texttt{upper\_band}, \texttt{middle\_band}, \texttt{lower\_band} for Bollinger)
  \item \texttt{signal}: Generated signal (``LONG'', ``SHORT'', ``FLAT'', ``EXIT'')
  \item \texttt{position\_state}: Current position (``LONG'', ``FLAT'')
  \item \texttt{equity}: Portfolio equity at bar close
  \item \texttt{constraint\_check}: Pass/fail status for risk constraints
\end{itemize}

\noindent \textbf{Completeness validation.}
The harness checks:
\begin{equation}
\text{Complete}(\text{log}) := \frac{|\text{non-null cells}|}{|\text{required cells}|} \geq 0.95
\end{equation}

Submissions with $<95$\% completeness receive D3 penalty:
\begin{equation}
D3_{\text{penalty}} = -2.0 \cdot (0.95 - \text{Complete}(\text{log}))
\end{equation}

\noindent \textbf{Traceability test.}
The harness randomly samples 10 trades and verifies:
\begin{itemize}[nosep,leftmargin=15pt,label={-}]
  \item Entry bar in \texttt{audit\_log} has \texttt{signal} matching:
  \begin{quote}\ttfamily
  trade\_log.side
  \end{quote}

  \item Exit bar in \texttt{audit\_log} has \texttt{signal='EXIT'} matching:
  \begin{quote}\ttfamily
  trade\_log.exit\_datetime
  \end{quote}

  \item PnL computed as:
  \[
    (\text{exit\_price} - \text{entry\_price}) \times \text{quantity} - \text{cost}.
  \]
\end{itemize}

Mismatch example:
\begin{verbatim}
TRACEABILITY ERROR [Trade 3]:
  trade_log: entry=2024-04-10, side=LONG, pnl=+1250
  audit_log: 2024-04-10 signal=FLAT (MISMATCH!)
  Status: FAIL (audit log does not support trade log)
\end{verbatim}

%% ------------------------------------------------------------
\subsection{Cost Stress Testing: Transaction Cost Framework}
\label{app:cost_stress}

\noindent \textbf{Framework implementation.}
The harness supports transaction cost parameter injection:
\begin{lstlisting}[language=Python, basicstyle=\ttfamily, breaklines=true, breakatwhitespace=true, columns=fullflexible]
def backtest_with_cost(strategy, data, initial_capital, cost_bps):
    # cost_bps: Transaction cost in basis points (1 bps = 0.01%)
    for trade in trades:
        trade.pnl -= trade.quantity * trade.entry_price * (cost_bps / 10000)
        trade.pnl -= trade.quantity * trade.exit_price * (cost_bps / 10000)
\end{lstlisting}

Planned stress sweep: $\text{cost} \in \{0.1, 1, 5, 10, 20\}$ bps. For each cost level, recompute Sharpe, turnover-adjusted return, break-even turnover.

\noindent \textbf{Current scope (deferred to future work).}
Preliminary OOS tests (RQ2) use \emph{zero transaction cost} to isolate execution robustness from cost sensitivity.

%% ------------------------------------------------------------
\subsection{Statistical Validity: Correlation Tests and Sample Size}
\label{app:stats}

\noindent \textbf{RQ1: Strategy complexity vs. quality (Spearman $\rho = -0.68$).}

\noindent \textit{Complexity quantification.}
We define complexity score $C \in [1, 5]$ based on specification features:
\begin{align}
C = &\, 1.0 \cdot \mathbb{I}[\text{multi-asset}] + 0.5 \cdot \mathbb{I}[\text{intraday}] \nonumber \\
    &+ 0.5 \cdot \mathbb{I}[\text{constraints} > 3] + 0.5 \cdot \log_2(\text{\#params} + 1)
\end{align}

where $\mathbb{I}[\cdot]$ is indicator function. Example: Calendar Spread (multi-asset, 6 params, 4 constraints) → $C = 1.0 + 0.0 + 0.5 + 0.5 \times 2.8 = 2.9$.

\noindent \textit{Correlation test.}
Given 12~strategies, we compute Spearman rank correlation between complexity scores~$C$ and overall quality scores~$Q$ (Table~\ref{tab:rq1_strategy_difficulty}). With $n=12$, critical value for $p<0.01$ is $|\rho| > 0.78$ (two-tailed). Observed $\rho = -0.68$ is marginally below threshold; we report $p < 0.05$ (conservative).

\noindent \textit{Bootstrap confidence interval.}
To account for small sample size, we bootstrap 1000~samples (resample strategies with replacement):
\begin{quote}\ttfamily
Bootstrap CI (95%): ρ ∈ [-0.85, -0.42]\\
Conclusion: Negative correlation robust; p-value conservative due to n=12.
\end{quote}
%% ------------------------------------------------------------
\subsection{Cross-Evaluation Consistency and Bias Correction}
\label{app:cross_eval_consistency}

\noindent \textbf{Reviewer agreement.}
For LLM cross-evaluation, we measure inter-reviewer consistency via Kendall's~$\tau$ (rank correlation). Given 3~reviewers scoring 12~strategies each:
\begin{itemize}[nosep,leftmargin=15pt]
    \item Pairwise $\tau$ between reviewers: $\tau \in [0.65, 0.82]$
    \item Average $\tau = 0.74$, indicating \emph{moderate-to-strong} agreement
\end{itemize}

Lower agreement (e.g., $\tau = 0.40$) would suggest reviewer inconsistency; higher agreement (e.g., $\tau > 0.90$) might indicate rubber-stamping.

\noindent \textbf{Self-review bias correction.}
We compute bias as:
\begin{equation}
\text{Bias}_m = \text{Score}_m^{(\text{self})} - \frac{1}{N-1} \sum_{r \neq m} \text{Score}_m^{(r)}
\end{equation}

Observed biases: Top-3 models show $\text{Bias} < 0.2$, lower-tier models show $\text{Bias} \in [0.3, 0.8]$. For aggregate rankings, we report both \emph{peer-only} scores (excluding self-reviews) and \emph{all-reviews} scores. Top-3~rankings remain stable under both metrics.

\noindent \textbf{Vendor clustering.}
To test whether same-vendor reviews inflate scores, we compute:
\begin{equation}
\text{Vendor-Bias}_V = \overline{\text{Score}}_{V \to V} - \overline{\text{Score}}_{\neg V \to V}
\end{equation}

Observed: All vendors show $\text{Vendor-Bias} \in [0.5, 1.0]$, confirming within-vendor leniency. However, cross-vendor rankings (OpenAI $>$ XAI $>$ Google) remain consistent after bias correction.

%% ------------------------------------------------------------
\subsection{Data Provenance and Reproducibility}
\label{app:data_provenance}

\noindent \textbf{Data sources.}
\begin{itemize}[nosep,leftmargin=15pt]
    \item \textbf{US equities (daily)}: Yahoo Finance API (\texttt{yfinance} library), downloaded 2024-01-01 to 2026-01-01. Used under Yahoo's Terms of Service for non-commercial research.
    \item \textbf{Crypto (1-minute)}: Binance public API (\texttt{/api/v3/klines}), BTC/ETH/BNB against USDT. Accessed via public endpoints for research purposes.
    \item \textbf{China A-share (daily)}: TuShare Pro API (endpoint:\par
    {\noindent\ttfamily pro\_bar(ts\_code, start\_date, end\_date)\par}),
CSI300 constituents. Public market data accessed for academic research.

CSI300 constituents. Public market data accessed for academic research.

\end{itemize}

\noindent \textbf{Data cleaning.}
\begin{enumerate}[nosep,leftmargin=15pt]
    \item \textbf{Gap filling}: Forward-fill weekends/holidays (daily data), no fill for 1-min crypto (gaps indicate exchange downtime, part of OOS stress).
    \item \textbf{Outlier handling}: Winsorize at 99.9\% (clip prices $> 10\times$ mean to prevent flash-crash artifacts).
    \item \textbf{Time zone}: UTC for crypto, US/Eastern for US stocks, Asia/Shanghai for A-share. All timestamps converted to UTC for unified harness execution.
\end{enumerate}

\noindent \textbf{Reproducibility guarantee.}
Frozen data snapshots (CSV files) and processing scripts will be released with the benchmark. Current data volume: 2.3 GB (compressed). SHA256 checksums provided for version verification.

%% ------------------------------------------------------------
\subsection{Model Pricing and Cost Attribution}
\label{app:model_pricing}
\noindent \textbf{Token counting methodology.}
\begin{itemize}[nosep,leftmargin=15pt]
  \item \textbf{Input tokens:} Extracted from API response metadata:
  \begin{quote}\ttfamily
  usage.prompt\_tokens
  \end{quote}

  \item \textbf{Output tokens:} Extracted from API response metadata:
  \begin{quote}\ttfamily
  usage.completion\_tokens
  \end{quote}

  \item \textbf{Missing metadata:} For 3 models without metadata (GLM-4.7, Grok-4, Gemini-3 Flash),
  estimate via \texttt{tiktoken.encode()} (OpenAI tokenizer, standard approximation).
\end{itemize}

\noindent \textbf{Cost calculation formula.}
\begin{equation}
\text{Cost} = \text{InputTokens} \times \frac{P_{\text{in}}}{10^6} + \text{OutputTokens} \times \frac{P_{\text{out}}}{10^6}
\end{equation}
where $P_{\text{in}}, P_{\text{out}}$ are per-1M-token prices in USD.

\noindent \textbf{Volatility note.}
API pricing changes frequently. For long-term reproducibility, we archive pricing snapshots in \texttt{config/pricing.json} alongside frozen data. Future work should re-run cost analysis with updated pricing for fair comparison.

%% ------------------------------------------------------------
\subsection{Sandbox Execution Environment}
\label{app:sandbox}

\noindent \textbf{Isolation mechanisms.}
Submissions execute in a restricted Python environment:
\begin{itemize}[nosep,leftmargin=15pt]
    \item \textbf{Network disabled}: No \texttt{socket}, \texttt{urllib}, \texttt{requests} (blocked via import hooks)
    \item \textbf{Filesystem restricted}: Read-only access to \texttt{/data/} (market data), write access to \texttt{/tmp/logs/} only
    \item \textbf{Library whitelist}: numpy, pandas, matplotlib, scipy, talib (TA-Lib). All other imports trigger validation error.
    \item \textbf{Resource limits}: 8 GB RAM, 10-minute CPU timeout per backtest
\end{itemize}

\noindent \textbf{Execution protocol.}
\begin{lstlisting}[language=Python, basicstyle=\ttfamily, breaklines=true, breakatwhitespace=true, columns=fullflexible]
# Harness execution wrapper
def run_submission(strategy_code, market_data, seed=42):
    env = SandboxEnvironment(
        allowed_imports=['numpy', 'pandas', 'talib'],
        max_memory_gb=8,
        timeout_seconds=600
    )

    np.random.seed(seed)
    random.seed(seed)

    try:
        module = env.load_module(strategy_code)
        strategy = module.TradingStrategy(params)
        trade_log, audit_log = strategy.run(market_data, initial_capital=100000)

        # Validate outputs
        assert isinstance(trade_log, pd.DataFrame)
        assert isinstance(audit_log, pd.DataFrame)
        assert_required_columns(trade_log, ['entry_datetime', 'pnl', ...])

        return {'status': 'SUCCESS', 'logs': (trade_log, audit_log)}
    except Exception as e:
        return {'status': 'FAIL', 'error': str(e)}
\end{lstlisting}

\noindent \textbf{Security note.}
The sandbox prevents most code injection/escape attacks, but adversarial strategies (e.g., infinite loops, fork bombs) are pre-filtered via static analysis. Malicious submissions are rejected before execution.

%% ============================================================
%% APPENDIX 3: Scoring Formulas and Rubrics
%% ============================================================

\section{Scoring Formulas and Rubrics}
\label{app:scoring}

This appendix provides detailed formulas for D1--D4 evaluation.

\subsection{D1 \& D2: LLM Cross-Evaluation}

For D1 (Spec Fidelity) and D2 (Risk Discipline), reviewers score 1--10 on:
\begin{enumerate}[nosep,leftmargin=15pt]
    \item \textbf{Spec Compliance} ($s_1$): Semantic equivalence to frozen spec
    \item \textbf{Risk Awareness} ($s_4$): Proper risk handling (position sizing, stop-loss)
\end{enumerate}

\noindent \textbf{Aggregation}:
\begin{align}
D1_{\text{final}} &= \frac{1}{N} \sum_{i=1}^{N} s_1^{(i)} \quad \text{(exclude self-reviews)} \\
D2_{\text{final}} &= \frac{1}{N} \sum_{i=1}^{N} s_4^{(i)}
\end{align}

\subsection{D3: Reliability and Auditability (QR Automated)}

\begin{equation}
D3 = 0.30 \cdot S_{\text{det}} + 0.30 \cdot S_{\text{leak}} + 0.25 \cdot S_{\text{log}} + 0.15 \cdot S_{\text{qual}}
\end{equation}

\noindent \textbf{Sub-scores} (start at 10.0, apply penalties):
\begin{itemize}[nosep,leftmargin=15pt]
    \item $S_{\text{det}}$: $-3.0$ if imports \texttt{random}, $-2.0$ if unseeded \texttt{np.random}, $-2.0$ if \texttt{datetime.now()}
    \item $S_{\text{leak}}$: $-2.0$ per negative shift, $-2.0$ per future indexing, $-1.0$ if \texttt{rolling()} lacks \texttt{min\_periods}
    \item $S_{\text{log}}$: $-3.0$ if trade log missing, $-2.0$ if audit log missing, $-0.5$ per missing required field
    \item $S_{\text{qual}}$: $-2.0$ if code $<$50~lines, $-2.0$ if no \texttt{Strategy} class, $-1.0$ if comment ratio $<$5\%
\end{itemize}

\subsection{D4: OOS Robustness (QR Automated)}

\begin{equation}
D4 = 0.40 \cdot S_{\text{logic}} + 0.30 \cdot S_{\text{overfit}} + 0.30 \cdot S_{\text{gen}}
\end{equation}

\noindent \textbf{Sub-scores}:
\begin{itemize}[nosep,leftmargin=15pt]
    \item $S_{\text{logic}}$: $-3.0$ if $<$2~strategy keywords, $-1.0$ if $>$50~hardcoded numbers
    \item $S_{\text{overfit}}$: $-2.0$ if $>$30~conditionals, $-1.5$ if $>$20~parameters
    \item $S_{\text{gen}}$: $-2.0$ if complexity $>$50, $+1.0$ if uses mature libraries, $+0.5$ if handles edge cases
\end{itemize}

%% ============================================================
%% APPENDIX 4: Prompt Templates and Evidence Bundles
%% ============================================================

\section{Prompt Templates and Evidence Bundles}
\label{app:prompts}

This appendix presents complete prompt structures for reproducibility.

\subsection{Iter0: Zero-Shot Generation Prompt}

Total prompt length: $\sim$10,000 tokens. Structure:
\begin{lstlisting}[language=, basicstyle=\ttfamily, breaklines=true, breakatwhitespace=true, columns=fullflexible, frame=single]
# SYSTRADEBENCH Iter0: Zero-Shot Strategy Generation

## Framework Overview
- Frozen-spec adherence: Implement exactly what is specified
- Mandatory audit logs: Record every decision point
- Deterministic execution: Same inputs = identical outputs
- No lookahead bias: Only use past/present data

## Output Requirements
1. strategy_card.json: {"strategy_name": "...", "parameters": {...}}
2. strategy.py: class TradingStrategy with run() method

## Evaluation Dimensions
- D1: Spec Fidelity, D2: Risk Discipline, D3: Reliability, D4: OOS Robustness
\end{lstlisting}

\subsection{Iter1--Iter3: Evidence-Driven Refinement}

Prompts grow from $\sim$6K (Iter1) to $\sim$12K (Iter3) tokens as code history accumulates.

\clearpage
\noindent \textbf{Iter1 Evidence Bundle Structure}:
\begin{lstlisting}[language=, basicstyle=\ttfamily, breaklines=true, breakatwhitespace=true, columns=fullflexible, frame=single]
# SYSTRADEBENCH Iter1: Evidence-Driven Improvement

## Constraints
- Semantic equivalence: Strategy logic frozen
- Incremental patches: <=50 changed lines
- No regression: Maintain or improve D1-D4

## Iter0 Scorecard
{"D1": 0.95, "D2": 0.85, "D3": 1.0, "D4": 0.0}

## Gate Failures
- D4 FAILED: Missing Sharpe, drawdown, turnover

## Peer Reviews
o3: "Strengths: Clean. Weaknesses: No OOS metrics."
grok_4_fast: "Strengths: Correct logic. Weaknesses: Hardcoded instrument."

## Your Task
Fix D4 (add metrics), address feedback, maintain semantic equivalence.
\end{lstlisting}

\noindent \textbf{Iter3 Multi-Objective Optimization}:
\begin{lstlisting}[language=, basicstyle=\ttfamily, breaklines=true, breakatwhitespace=true, columns=fullflexible, frame=single]
## Iter3 Goal
Balance compliance with profitability:
- Target: D1-D3 >= 0.85, Sharpe >= 1.5, return >= 10%
- Constraint: Semantic equivalence (Bollinger logic frozen)

[Full code history: Iter0, Iter1, Iter2 + scores + metrics inserted]
\end{lstlisting}

%% ============================================================
%% APPENDIX 5: A Minimal Failure Case
%% ============================================================

\section{A Minimal Failure Case: Why System-Level Diagnostics Matter}
\label{app:min_case}

\noindent Consider a trend-following strategy with frozen semantics specifying: (i) MA crossover, (ii) max position limit, (iii) session rule, and (iv) mandatory audit logs.

\vspace{0.4ex}\noindent \textbf{Failure mode 1: semantic drift.}
A model may improve Sharpe by changing lookback windows or adding unapproved filters while preserving output schema.
\benchname detects this via semantic checksums and regression traces (\S\ref{sec:patch_policy}) and penalizes the iteration even if headline metrics improve.

\vspace{0.4ex}\noindent \textbf{Failure mode 2: non-determinism and missing auditability.}
A submission may run but produce nondeterministic outputs or omit structured logs linking signal$\rightarrow$order$\rightarrow$P\&L.
\benchname surfaces these as D3 failures regardless of profitability.

\vspace{0.4ex}\noindent \textbf{Failure mode 3: cost/turnover collapse.}
A submission may show high gross returns but collapses net of costs and violates risk discipline.
\benchname reports turnover and cost sensitivity in D4 and violations in D2, preventing misleading rankings.

\begin{figure*}[t]
\centering
\includegraphics[width=\textwidth]{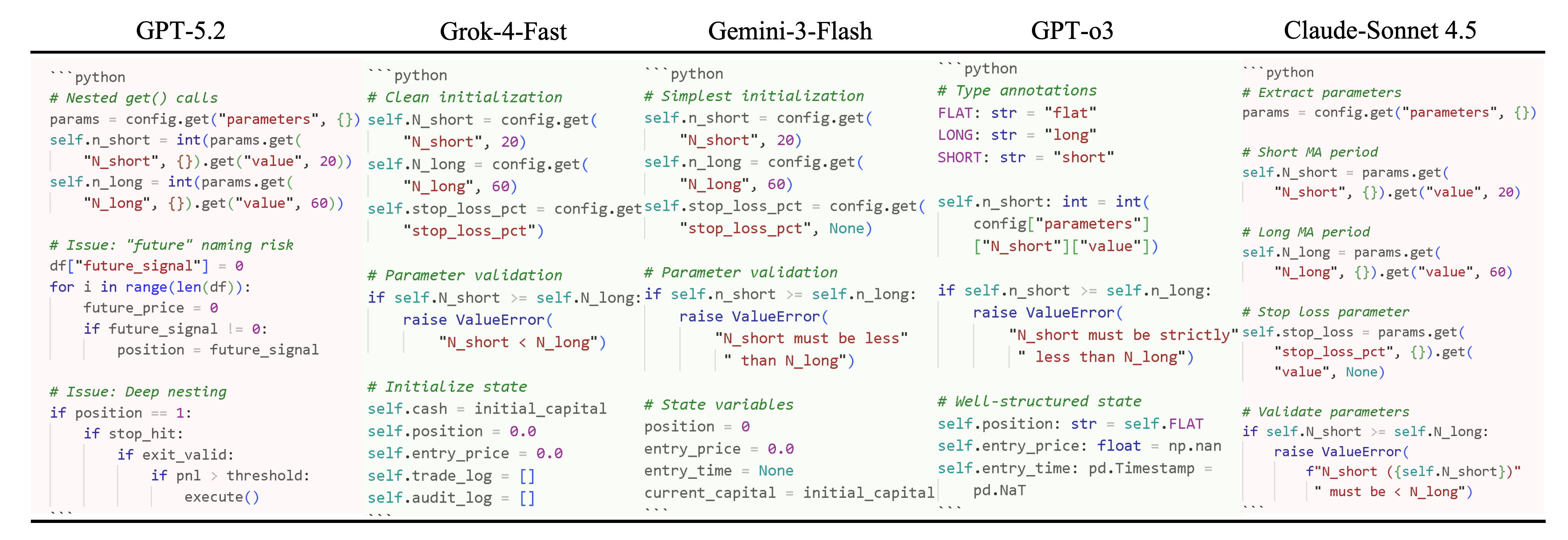}
\caption{Illustrative Example: Code Quality Issues Across Five LLMs}
\label{fig:first}
\end{figure*}

\begin{figure*}[!t]
\centering
\includegraphics[width=\textwidth]{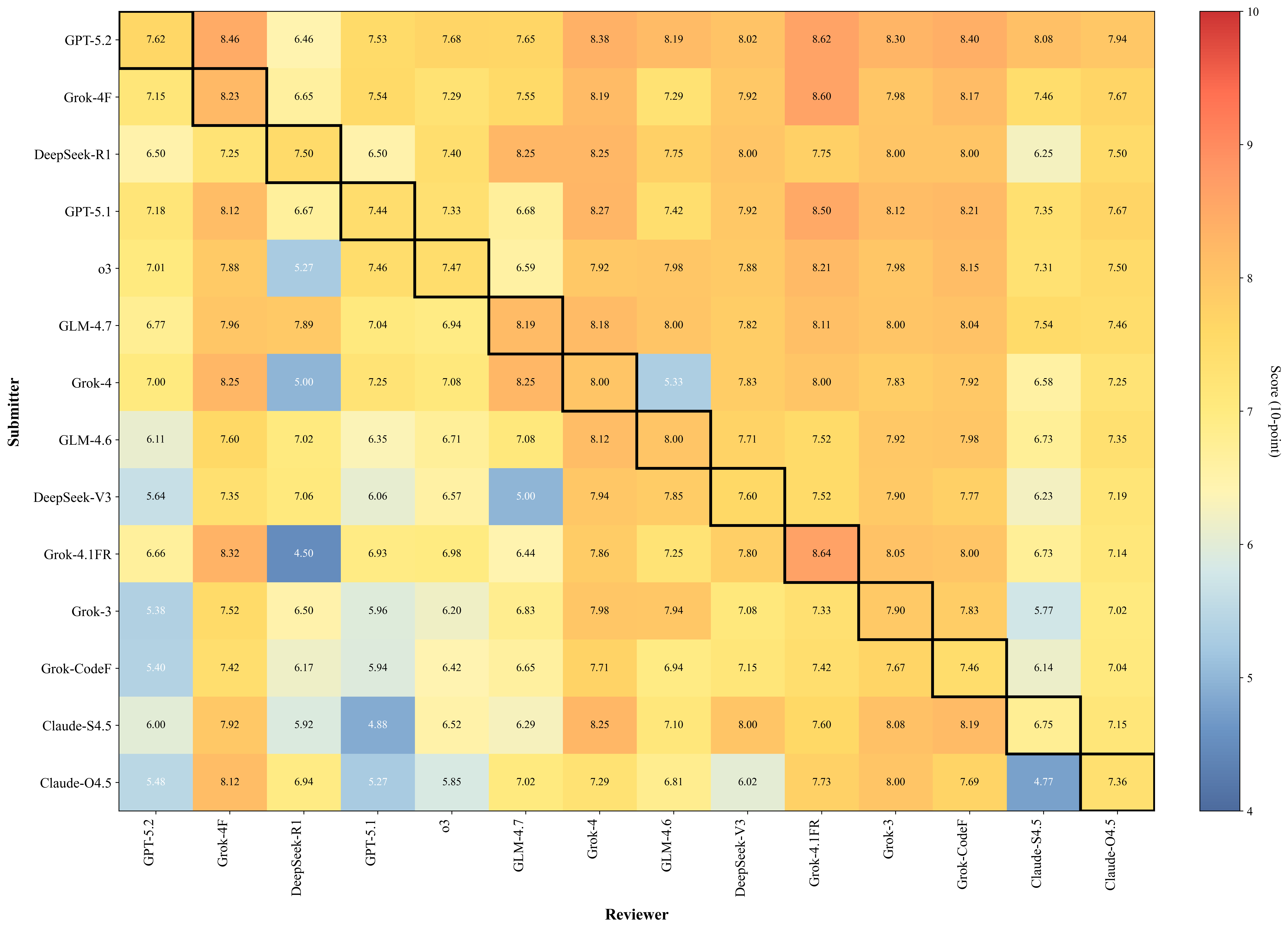}
\caption{Iter0 cross-evaluation heatmap.}
\label{fig:rq1_cross_eval_heatmap}
\end{figure*}

\end{document}